\documentclass[12pt]{article}
\makeatletter\renewcommand{\section}{\@startsection
{section}{1}{\z@}{-3.5ex plus -1ex minus
    -.2ex}{2.3ex plus .2ex}{\bf }}

\makeatletter\renewcommand{\subsection}{\@startsection{subsection}{2}{\z@}{-3.25ex plus -1ex minus
   -.2ex}{1.5ex plus .2ex}{\it }}
\makeatletter\renewcommand{\subsubsection}{\@startsection{subsubsection}{3}{-2.45ex}{-3.25ex
plus -1ex minus -.2ex}{1.5ex plus .2ex}{\it }}

\let\fn\footnote
\renewcommand{\footnote}[1]{\linespread{1.1}\fn{#1}\linespread{1.29}}

\usepackage[english]{babel}
\usepackage{amsmath,amssymb}
\usepackage{bbm}

\def\slasha#1{\setbox0=\hbox{$#1$}#1\hskip-\wd0\hbox to\wd0{\hss\sl/\/\hss}}

\def\periodb#1{\setbox0=\hbox{$#1$}#1\hskip-\wd0\hbox to\wd0{-}}
\newcommand{\unit}{\mathbbm{1}}   
\newcommand{\CA}{\mathcal{A}}    
\newcommand{\CD}{\mathcal{D}}    
\newcommand{\CF}{\mathcal{F}}    
\newcommand{\CI}{\mathcal{I}}    
\newcommand{\CK}{\mathcal{K}}    
\newcommand{\CM}{\mathcal{M}}    
\newcommand{\CN}{\mathcal{N}}    
\newcommand{\CO}{\mathcal{O}}    
\newcommand{\CP}{\mathcal{P}}    
\newcommand{\CU}{\mathcal{U}}    
\newcommand{\CE}{\mathcal{E}}    
\newcommand{\frA}{\mathfrak{A}}    
\newcommand{\frG}{\mathfrak{G}}    
\newcommand{\frH}{\mathfrak{H}}    
\newcommand{\frS}{\mathfrak{S}}    
\newcommand{\frU}{\mathfrak{U}}    
\newcommand{\frX}{\mathfrak{X}}    
\newcommand{\FR}{\mathbbm{R}}     
\newcommand{\FC}{\mathbbm{C}}     
\newcommand{\CPP}{{\mathbbm{C}P}}    
\newcommand{\RZ}{\mathbbm{Z}}     
\newcommand{\dd}{\mathrm{d}}     
\newcommand{\dpar}{\partial}     
\newcommand{\dparb}{{\bar{\partial}}}     
\newcommand{\embd}{{\hookrightarrow}}     
\newcommand{\bz}{{\bar{z}}}     
\newcommand{\bl}{{\bar{\lambda}}}     
\newcommand{\ald}{{\dot{\alpha}}}     
\newcommand{\bed}{{\dot{\beta}}}     
\newcommand{\gad}{{\dot{\gamma}}}     
\newcommand{\ded}{{\dot{\delta}}}     
\newcommand{\eps}{{\varepsilon}}     
\newcommand{\eand}{{~~~\mbox{and}~~~}}     
\newcommand{\kernel}{{\mathrm{ker}}}     
\newcommand{\etr}{{\mathrm{etr}}}     
\newcommand{\edet}{{\mathrm{edet}}}     
\newcommand{\der}[1]{\frac{\dpar}{\dpar #1}}   
\newcommand{\tr}{\,\mathrm{tr}\,}     

\newcommand{\remark}[1]{}     

\newcommand{\SA}{\CP^{3\oplus 2|0}}     
\newcommand{\SB}{\CP^{3\oplus 1|0}}     

\makeatletter
\@addtoreset{equation}{section}
\makeatother

\begin{document}
\begin{titlepage}
\setcounter{page}{0}
\begin{flushright}
hep--th/0410292\\
ITP--UH--26/04
\end{flushright}
\vskip 1.5cm
\begin{center}
{\LARGE \bf The Topological B-Model on Fattened\\ Complex Manifolds and Subsectors of \\[0.5cm]
$\CN=4$ Self-Dual Yang-Mills Theory} \vskip 1.5cm
\renewcommand{\thefootnote}{\fnsymbol{footnote}}
{\Large Christian S\"{a}mann} \setcounter{footnote}{0}
\renewcommand{\thefootnote}{\arabic{thefootnote}}
\vskip 1cm
{\em Institut f\"{u}r Theoretische Physik\\
Universit\"{a}t Hannover\\
Appelstra{\ss}e 2, 30167 Hannover, Germany}\\[5mm]
{Email: saemann@itp.uni-hannover.de} \vskip 1.1cm
\end{center}
\begin{center}
{\bf Abstract}
\end{center}
\begin{quote}
In this paper, we propose so-called fattened complex manifolds as
target spaces for the topological B-model. We naturally obtain
these manifolds by restricting the structure sheaf of the $\CN=4$
supertwistor space, a process, which can be understood as a
fermionic dimensional reduction. Using the twistorial description
of these fattened complex manifolds, we construct Penrose-Ward
transforms between solutions to the holomorphic Chern-Simons
equations on these spaces and bosonic subsectors of solutions to
the $\CN=4$ self-dual Yang-Mills equations on $\FC^4$ or $\FR^4$.
Furthermore, we comment on Yau's theorem for these spaces.
 \vskip 5mm
\end{quote}
\end{titlepage}
\newpage
\setcounter{page}{1}

\section{Introduction}

Calabi-Yau manifolds play an important r{\^o}le in topological string
theory. Besides other interesting features as e.g.\ mirror
symmetry, they are suited as target spaces for the so-called
topological B-model. This B-model has been shown to be equivalent
to holomorphic Chern-Simons (hCS) theory defined on its target
space \cite{Witten:1992fb}.

In \cite{Witten:2003nn}, Witten studied the supermanifold
$\CPP^{3|4}$ which is a Calabi-Yau supermanifold and
simultaneously a supertwistor space. For this space, the
equivalence of the topological B-model with hCS theory still holds
and the moduli space of classical solutions of hCS theory can be
bijectively mapped to the moduli space of solutions to the $\CN=4$
supersymmetrically extended self-dual Yang-Mills (SDYM) equations
via a Penrose-Ward correspondence\footnote{For reviews of twistor
theory and the Penrose-Ward correspondence, see
\cite{twistorbooks}.} \cite{Witten:2003nn} (for the discussion
beyond linearized level, see \cite{Popov:2004rb}). A crucially new
aspect of this supertwistor description is the possibility of
giving an action for hCS theory on the Calabi-Yau supermanifold,
which is due to the existence of a holomorphic volume form on that
space. This has never been achieved on the purely bosonic twistor
space.

The link between topological string theory and SDYM theory was
subsequently used to calculate gauge theory amplitudes with string
theory machinery and other methods \cite{amp}-\cite{motl}; even
gravity amplitudes have been considered \cite{grav}. A variety of
further issues appearing in the context of twistor strings has
also been examined, see \cite{Sinkovics:2004fm} or
\cite{Kulaxizi:2004pa} for recent examples.

With the extensive use of Calabi-Yau supermanifolds, the question
of how mirror symmetry fits into the picture became more and more
interesting \cite{mirror}, \cite{Ahn:2004xs}. Particularly for the
discussion of this point, it is useful to have more spaces at hand
which bring along the helpful feature of a twistorial description.

Besides the Calabi-Yau supermanifold $\CPP^{3|4}$ introduced in
\cite{Witten:2003nn} and the weighted projective superspaces
proposed already in \cite{Witten:2003nn} and considered in
\cite{Popov:2004nk}, \cite{Ahn:2004xs} and \cite{Ahn:2004yu},
there is another class of presumably interesting spaces in direct
reach: the so-called ``fattened complex manifolds''
\cite{Eastwood:1992}. These manifolds are extensions of ordinary
manifolds with additional dimensions described by even nilpotent
coordinates. Similar spaces have been studied in the mathematical
literature since the early 1960s \cite{Grauert}. A class of
examples of them can formally be obtained by pairing the Gra{\ss}mann
coordinates of a supermanifold. But fattened complex manifolds
even allow for more constructive freedom: One can define a
twistorial Calabi-Yau supermanifold corresponding to a fattened
complex manifold with additional fermionic extensions. On the
fattened complex manifolds which we will study in the following,
hCS theory corresponds to bosonic subsectors of $\CN=4$ SDYM
theory similarly to the case of weighted projective spaces.
However, the parity of the truncated field content and thus also
the equations of motion will differ from some of the results
obtained in \cite{Popov:2004nk} for weighted projective spaces.

In this paper, we first discuss the existing framework for exotic
supermanifolds, i.e.\ generalized supermanifolds which have
additional even nilpotent coordinates. We then briefly review
supertwistor geometry before constructing hCS and SDYM theory for
two cases of exotic supermanifolds via a twistor correspondence.
In the real case, we give a field expansion for the gauge
potential of hCS theory, making the relation to SDYM theory
explicit. We close with some remarks on the extension of Yau's
theorem to exotic supermanifolds which are Calabi-Yau. This
theorem guarantees the existence of a Ricci-flat metric in every
K\"{a}hler class for K\"{a}hler manifolds with vanishing first Chern
class. The results we find are very similar to
\cite{Rocek:2004bi}: The straightforward extension of Yau's
theorem to exotic supermanifolds is not valid in general, but an
additional constraint has to be imposed.

\section{Exotic supermanifolds}

In this section, we want to give a brief review of the existing
extensions or generalizations of supermanifolds, which are suited
as a general framework for describing the target spaces for the
B-model employed later on. In the following, we call every (in a
well-defined way generalized) manifold which is locally described
by $k$ even, $l$ even and nilpotent and $q$ odd and nilpotent
coordinates an {\em exotic supermanifold} of dimension $(k\oplus
l|q)$. For a review on Gra{\ss}mann algebras, supernumbers and
supermanifolds, see e.g.\ \cite{Cartier:2002zp}.

\subsection{Partially formal supermanifolds}

The objects of supermathematics, as e.g.\ supermanifolds or
supergroups, are naturally described as covariant functors from
the category of Gra{\ss}mann algebras to corresponding categories of
ordinary mathematical objects, as manifolds or groups,
\cite{Schwarz:1984}. A generalization of this setting is to
consider covariant functors with the category of {\em almost
nilpotent} (AN) {\em algebras}\/\footnote{An {\em almost nilpotent
algebra} is an associative, finite-dimensional, unital,
$\RZ_2$-graded supercommutative algebra in which the ideal of
nilpotent elements has codimension 1.} as domain
\cite{Konechny:1997hr}. Recall that an AN algebra $\Xi$ can be
decomposed into an even part $\Xi_0$ and an odd part $\Xi_1$ as
well as in the canonically embedded ground field (i.e.\ $\FR$ or
$\FC$), $\Xi_B$, and the nilpotent part $\Xi_S$. The parts of
elements $\xi\in\Xi$ belonging to $\Xi_B$ and $\Xi_S$ are called
the {\em body} and the {\em soul} of $\xi$, respectively.

A {\em superspace} is a covariant functor from the category of AN
algebras to the category of sets. Furthermore, a {\em topological
superspace} is a functor from the category of AN algebras to the
category of topological spaces.

Consider now a tuple
$(x^1,...,x^k,y^1,...,y^l,\zeta^1,...,\zeta^q)$ of $k$ even, $l$
even and nilpotent and $q$ odd and nilpotent elements of an AN
algebra $\Xi$, i.e.\ $x^i\in\Xi_0$, $y^i\in\Xi_0\cap\Xi_S$ and
$\zeta^i\in\Xi_1$. The functor from the category of AN algebras to
such tuples is a superspace denoted by $\FR^{k\oplus l|q}$. An
open subset $U^{k\oplus l|q}$ of $\FR^{k\oplus l|q}$, which is
obtained by restricting the fixed ground field $\Xi_B$ of the
category of AN algebras to an open subset, is called a {\em
superdomain} of dimension $(k\oplus l|q)$. After defining a graded
basis $(e_1,...,e_k,f_1,...,f_{l},\eps_1,...,\eps_q)$ consisting
of $k+l$ even and $q$ odd vectors, one can consider the set of
linear combinations
$\left\{x^ie_i+y^jf_j+\zeta^\alpha\eps_\alpha\right\}$ which forms
a {\em supervector space} \cite{Konechny:1997hr}.

Roughly speaking, one defines a partially formal
supermanifold\footnote{This term was introduced in
\cite{Kontsevich:1997}.} of dimensions $(k\oplus l|q)$ as a
topological superspace smoothly glued together from superdomains
$U^{k\oplus l|q}$. Although we will not need the exact definition
in the subsequent discussion, we will nevertheless give it here
for completeness sake.

We define a {\em map between two superspaces} as a natural
transformation of functors. More explicitly, consider two
superspaces $\CM$ and $\CN$. Then a map $F:\CM\rightarrow\CN$ is a
map between superspaces, if $F$ is compatible with the morphisms
of AN algebras $\alpha:\Xi\rightarrow\Xi'$. We call a smooth map
$\kappa:\FR^{k\oplus l|q}_\Xi\rightarrow\FR^{k'\oplus l'|q'}_\Xi$
between two superdomains $\Xi_0$-{\em smooth}, if for every
$x\in\FR^{k\oplus l|q}_\Xi$ the tangent map
$(\kappa_\Xi)_*:T_x\rightarrow T_{\kappa_\Xi(x)}$ is a
homomorphism of $\Xi_0$-modules. Furthermore, we call a map
$\kappa:\FR^{k\oplus l|q}\rightarrow\FR^{k'\oplus l'|q'}$ {\em
smooth}, if for all AN algebras $\Xi$ the maps $\kappa_\Xi$ are
$\Xi_0$-smooth.

Now we can be more precise: A {\em partially formal supermanifold
of dimension} $(k\oplus l|q)$ is a superspace locally equivalent
to superdomains of dimension $(k\oplus l|q)$ with smooth
transition functions on the overlaps. Thus, a partially formal
supermanifold is also an exotic supermanifold.

However, not every exotic supermanifold is partially formal. We
will shortly encounter examples of such cases: exotic
supermanifolds, which are constructed using a particular AN
algebra instead of working with the category of AN algebras.

The definitions used in this section stem from
\cite{Konechny:1997hr}, where one also finds examples of
applications.

Unfortunately, it is not clear how to define a general integration
over the even nilpotent part of such spaces; even the existence of
such an integral is questionable. We will comment on this point
later on. As we need an integration to define an action for our
models, we have to turn to other generalizations.

\subsection{Thick complex manifolds}

Extensions to $m$-th formal neighborhoods of a submanifold $X$ in
a manifold $Y\supset X$ and the more general thickening procedure
have been proposed and considered long ago\footnote{In fact, the
study of infinitesimal neighborhoods goes back to \cite{Grauert}
and \cite{Griffiths}. For a recent review, see
\cite{Camacho:2002}.} in the context of twistor theory, in
particular for ambitwistor spaces, e.g.\ in
\cite{Witten:1978xx,Eastwood:1987,LeBrun:1986,Eastwood:1986}. We
will ignore this motivation and only recollect the definitions
needed for our subsequent discussion.

Given a complex manifold $X$ with structure sheaf $\CO_X$, we
consider a sheaf of $\FC$-algebras $\CO_{(m)}$ on $X$ with a
homomorphism $\alpha:\CO_{(m)}\rightarrow\CO_X$, such that locally
$\CO_{(m)}$ is isomorphic to $\CO[y]/(y^{m+1})$ where $y$ is a
formal (complex) variable and $\alpha$ is the obvious projection.
The resulting ringed space $X_{(m)}:=(X,\CO_{(m)})$ is called a
{\em thick complex manifold}. Similarly to the nomenclature of
supermanifolds, we call the complex manifold $X$ the {\em body} of
$X_{(m)}$.

As a simple example, let $X$ be a closed submanifold of the
complex manifold $Y$ with codimension one. Let $\CI$ be the ideal
of functions vanishing on $X$. Then $\CO_{(m)}=\CO_Y/\CI^{m+1}$ is
called an {\em infinitesimal neighborhood} or the $m$-th {\em
formal neighborhood} of $X$. This is a special case of a thick
complex manifold. Assuming that $X$ has complex dimension $n$,
$\CO_{(m)}$ is also an exotic supermanifold of dimension $(n\oplus
1|0)$. More explicitly, let $(x^1,...,x^n)$ be local coordinates
on $X$ and $(x^1,...,x^n,y)$ local coordinates on $Y$. Then the
ideal $\CI$ is generated by $y$ and $\CO_{(m)}$ is locally a
formal polynomial in $y$ with coefficients in $\CO_X$ together
with the identification $y^{m+1}\sim 0$. Furthermore, one has
$\CO_{(0)}=\CO_X$.

Returning to the local description as a formal polynomial in $y$,
we note that there is no object $y^{-1}$ as it would violate
associativity by an argument like $0=y^{-1} y^{m+1}=y^{-1}y
y^m=y^m$. However, the inverse of a formal polynomial in $y$ is
defined if (and only if) the zeroth order monomial has an inverse.
Suppose $p=a+\sum_{i=1}^mf_i y^i=a+b$, then we have
$p^{-1}=\frac{1}{a}\sum_{i=0}^m(-\frac{b}{a})^i$, analogously to
the inverse of a supernumber.

A {\em holomorphic vector bundle} on $(X,\CO_{(m)})$ is a locally
free sheaf of $\CO_{(m)}$-modules.

The {\em tangent space} of a thick complex manifold is the sheaf
of derivations $D:\CO_{(m)}\rightarrow\CO_{(m)}$. Let us consider
again our above example $X_{(m)}=(X,\CO_{(m)})$. Locally, an
element of $T X_{(m)}$ will take the form $D=f
\der{y}+\sum_jg^j\der{x^j}$ together with the differentiation
rules
\begin{equation}
\der{y}\,y=1~,~~~\der{y}\,x^i=\der{x^i}\,y=0~,~~~\der{x^i}\,x^j=\delta_i^j~.
\end{equation}

All this and the introduction of cotangent spaces for thick
complex manifolds is found in \cite{Eastwood:1986}.

In defining a (definite) integral over the nilpotent formal
variable $y$, which is needed for formulating hCS theory by giving
an action, one faces the same difficulty as in the case of Berezin
integration: the integral should not be taken over a specific
range as we integrate over an infinitesimal neighborhood which
would give rise to infinitesimal intervals. Furthermore, this
neighborhood is purely formal and so has to be the integration.
Recall that a suitable integration $I$ should satisfy the
rule\footnote{This rule can also be used to fix Berezin
integration.} $DI=ID=0$, where $D$ is a derivative with respect to
a variable over which $I$ integrates. The first requirement $DI=0$
states that the result of definite integration does not depend on
the variables integrated over. The requirement $ID=0$ for
integration domains with vanishing boundary (or functions
vanishing on the boundary) is the foundation of Stokes' formula
and integration by parts. It is easy to see that the condition
$DI=ID=0$ demands that
\begin{equation}
I=c\cdot\frac{\dpar^m}{\dpar y^m}~,
\end{equation}
where $y$ is the local formal variable from the definition of
$X_{(m)}$ and $c$ is an arbitrary normalization constant, e.g.\
$c=1/m!$ would be most convenient. Thus, we define
\begin{equation}
\int \dd y f:=\frac{1}{m!}\,\frac{\dpar^m}{\dpar y^m}\, f~.
\end{equation}
This definition only relies on an already well-defined operation
and thus is well-defined itself.\footnote{From this definition, we
see the problem arising for partially formal supermanifolds: The
integration process on thick complex manifolds returns the
coefficient of the monomial with highest possible power in $y$.
For partially formal supermanifolds, where one works with the
category of AN algebras, such a highest power does not exist as it
is different for each individual AN algebra.} Additionally, it
also agrees with the intuitive picture, that the integral of a
constant over an infinitesimal neighborhood should vanish.
Integration over a thick complex manifold is an
integro-differential operation.

Consider now a change of coordinates $(x^1,...,x^n,y)\rightarrow
(\tilde{x}^1,...,\tilde{x}^n,\tilde{y})$ which leaves invariant
the structure of the thick complex manifold. That is,
$\tilde{x}^i$ is independent of $y$, and $\tilde{y}$ is a
polynomial only in $y$ with vanishing zeroth order coefficient and
non-vanishing first order coefficient. Because of
$\dpar_{\tilde{y}}=\frac{\dpar y}{\dpar \tilde{y}}\,\dpar_y$, we
have the following transformation of a volume element under such a
coordinate change:
\begin{equation}
\dd \tilde{x}^1...\dd \tilde{x}^n\dd
\tilde{y}=\det\left(\frac{\dpar \tilde{x}^i}{\dpar x^j}\right)\dd
x^1...\dd x^n\left(\frac{\dpar y}{\dpar \tilde{y}}\right)^m\dd y~.
\end{equation}

The theorems in \cite{Eastwood:1986} concerning obstructions to
finding $X_{(m+1)}$ given $X_{(m)}$ will not be needed in the
following, as we will mainly work with order one thickenings (or
fattenings) and in the remaining cases, the existence directly
follows by construction.

\subsection{Fattened complex manifolds}

Fattened complex manifolds \cite{Eastwood:1992} are
straightforward generalizations of thick complex manifolds.
Consider again a complex manifold $X$ with structure sheaf
$\CO_X$. The {\em $m$-th order fattening with codimension $k$} of
$X$ is the ringed space $X_{(m,k)}=(X,\CO_{(m,k)})$ where
$\CO_{(m,k)}$ is locally isomorphic to
$\CO[y^1,...,y^k]/(y^1,...,y^k)^{m+1}$. Here the $y^i$ are again
formal complex variables. We also demand the existence of the
(obvious) homomorphism $\alpha:\CO_{(m,k)}\rightarrow\CO_X$. It
follows immediately, that a fattening with codimension 1 is a
thickening. Furthermore, an $(m,k)$-fattening of an
$n$-dimensional complex manifold $X$ is an exotic supermanifold of
dimension $(n\oplus k|0)$ and we call $X$ the {\em body} of
$X_{(m,k)}$.

As in the case of thick complex manifolds, there are no inverses
for the $y^i$, but the inverse of a formal polynomial $p$ in the
$y^i$ decomposed into $p=a+b$, where $b$ is the nilpotent part of
$p$, exists again if and only if $a\neq 0$ and it is then given by
$p^{-1}=\frac{1}{a}\sum_{i=0}^m(-\frac{b}{a})^i$. A {\em
holomorphic vector bundle} on $\CO_{(m,k)}$ is a locally free
sheaf of $\CO_{(m,k)}$-modules. The {\em tangent space} of a thick
complex manifold is also generalized in an obvious manner.

We define the integral analogously to thick complex manifolds as
\begin{equation}
\int \dd y^1...\dd y^k f:=\frac{1}{m!}\,\frac{\dpar^m}{\dpar
(y^1)^m}~...~\frac{1}{m!}\,\frac{\dpar^m}{\dpar (y^k)^m}\, f~.
\end{equation}
A change of coordinates
$(x^1,...,x^n,y^1,...,y^k)\rightarrow(\tilde{x}^1,...,\tilde{x}^n,\tilde{y}^1,...,\tilde{y}^k)$
must again preserve the structure of the fat complex manifold:
$\tilde{x}^i$ is independent of the $y^i$ and the $\tilde{y}^i$
are nilpotent polynomials in the $y^i$ with vanishing monomial of
order 0 and at least one nonvanishing monomial of order 1.
Evidently, all the $\tilde{y}^i$ have to be linearly independent.
Such a coordinate transformation results in a more complicated
transformation law for the volume element:
\begin{equation}
\dd \tilde{x}^1...\dd \tilde{x}^n\dd
\tilde{y}^1...\dd\tilde{y}^k=\det\left(\frac{\dpar
\tilde{x}^i}{\dpar x^j}\right)\dd x^1...\dd x^n\left(\frac{\dpar
y^{i_1}}{\dpar \tilde{y}^1}...\frac{\dpar y^{i_k}}{\dpar
\tilde{y}^k}\right)^m\dd y^{i_1}...\dd y^{i_k}~,
\end{equation}
where a sum over the indices $(i_1,...,i_k)$ is implied. In this
case, the coefficient for the transformation of the nilpotent
formal variables cannot be simplified. Recall that in the case of
ordinary differential forms, the wedge product provides the
antisymmetry needed to form the determinant of the Jacobi matrix.
In the case of Berezin integration, the anticommutativity of the
derivatives with respect to Gra{\ss}mann variables does the same for
the inverse of the Jacobi matrix. Here, we have neither of these
and therefore no determinant appears.

\subsection{Thick and fattened supermanifolds}

After thickening or fattening a complex manifold, one can readily
add fermionic dimensions. Given a thickening of an $n$-dimensional
complex manifold of order $m$, the simplest example is possibly
$\Pi T X_{(m)}$, an $(n\oplus 1|n+1)$ dimensional exotic
supermanifold. However, we will not study such objects in the
following.

\subsection{Exotic Calabi-Yau supermanifolds}

It has become common usage to call a supermanifold with vanishing
first Chern class a Calabi-Yau supermanifold, even if not all such
spaces admit a Ricci-flat metric. Counterexamples to Yau's theorem
for Calabi-Yau supermanifolds can be found in \cite{Rocek:2004bi}.
Following this convention, we call an exotic supermanifold
Calabi-Yau, if its first Chern class vanishes and it therefore
comes with a holomorphic volume form. Also for exotic
supermanifolds, the Calabi-Yau property is not sufficient for the
existence of a Ricci-flat metric, as we will find in the last
section.

Nevertheless, one should remark that vanishing of the first Chern
class -- and not Ricci-flatness -- is necessary for a consistent
definition of the B-model on a manifold. And, from another
viewpoint, it is only with the help of a holomorphic volume form,
that one can give an action for hCS theory. Thus, the nomenclature
is justified from a physicist's point of view.

\subsection{Dolbeault and \v{C}ech descriptions of holomorphic vector
bundles}

The twistor correspondence \cite{twistorbooks} makes heavy use of
two different descriptions of holomorphic vector bundles: the
Dolbeault and the \v{C}ech description. Let us briefly comment on
the extension of both to fattened complex manifolds.

Consider a trivial principal $G$-bundle $P$ over a fattened
complex manifold $X$ covered by a collection of patches
$\frU=\{\CU_a\}$ with coordinates $(z^i_a,y^j_a)$ and let $G$ have
a representation in terms of $n\times n$ matrices.

Let $\frG$ be an arbitrary sheaf of $G$-valued functions on $X$.
The set of {\em \v{C}ech $q$-cochains} $C^q(\frU,\frG)$ is the
collection $\psi=\{\psi_{a_0...a_q}\}$ of sections of $\frG$
defined on nonempty intersections $\CU_{a_0}\cap...\cap
\CU_{a_q}$. Furthermore, we define the sets of \v{C}ech 0- and
1-cocycles by\footnote{Note that any regular matrix-valued
function on a fattened complex manifold of order $m$ can be
decomposed into an ordinary matrix-valued function and a nilpotent
matrix-valued function, both defined on the body of the fattened
complex manifold: $\psi=\psi_0+\psi_n$. The inverse of such a
function $\psi$ exists, if $\psi_0$ is invertible and then it is
explicitly given by
\begin{equation*}
\psi^{-1}=\psi_0^{-1}-\psi_0^{-1}\psi_n\psi_0^{-1}+
\psi_0^{-1}\psi_n\psi_0^{-1}\psi_n\psi_0^{-1}-
\psi_0^{-1}\psi_n\psi_0^{-1}\psi_n\psi_0^{-1}\psi_n\psi_0^{-1}+...~,
\end{equation*}
with $m+1$ terms altogether.}
\begin{align}
&Z^0(\frU,\frG):=\{\,\psi\in
C^0(\frU,\frG)~|~\psi_a=\psi_b~\mbox{on}~\CU_a\cap \CU_b\neq
\varnothing\}=\Gamma(\frU,\frG)~,\\\nonumber
&Z^1(\frU,\frS):=\{\,\chi\in
C^1(\frU,\frG)~|~\chi_{ab}=\chi^{-1}_{ba}~\mbox{on}~ \CU_a\cap
\CU_b\neq
\varnothing,~\\&\hspace{2.7cm}\chi_{ab}\chi_{bc}\chi_{ca}=\unit~\mbox{on}~
\CU_a\cap \CU_b\cap \CU_c\neq \varnothing\}~.
\end{align}
This definition implies, that the \v{C}ech 0-cocyles are
independent of the covering: $Z^0(\frU,\frG)=Z^0(X,\frG)$, and we
define the {\em zeroth \v{C}ech cohomology set} by
$H^0(X,\frG):=Z^0(X,\frG)$. Two 1-cocycles $\tilde{\chi}$ and
$\chi$ are called {\em equivalent} if there is a 0-cochain
$\psi\in C^0(\frU,\frG)$ such that
$\tilde{\chi}_{ab}=\psi_a\chi_{ab}\psi_b^{-1}$ on all
$\CU_a\cap\CU_b\neq \varnothing$. Dividing $Z^1(\frU,\frG)$ by
this equivalence relation gives the {\em first \v{C}ech cohomology
set} $H^1(\frU,\frG)\cong Z^1(\frU,\frG)/C^0(\frU,\frG)$.

Besides the \v{C}ech cohomology sets, we also need the sheaf
$\frS$ of smooth $G$-valued functions on $X$ and its subsheaf
$\frH$ of holomorphic $G$-valued functions on $X$. Furthermore, we
denote by $\frA$ the sheaf of flat (0,1)-connections on $P$, i.e.\
germs of solutions to
\begin{equation}\label{eomdol}
\dparb \CA^{0,1}+\CA^{0,1}\wedge \CA^{0,1}=0~,
\end{equation}
which are purely holomorphic in the fattened directions, i.e.\
\begin{equation}\label{eomdol2}
\der{\bar{y}^j_a}\,\CA_a^{0,1}=0~~~\mbox{on all}~~\CU_a~.
\end{equation}

Note that elements $\CA^{0,1}$ of $H^0(X,\frA)=\Gamma(X,\frA)$
define a holomorphic structure $\dparb_\CA=\dparb+\CA^{0,1}$ on a
trivial rank $n$ complex vector bundle over $X$. The moduli space
$\CM$ of such holomorphic structures is obtained by dividing
$H^0(X,\frA)$ by the group of gauge transformations, i.e.\
elements $g$ of $H^0(X,\frS)$ acting on elements $\CA^{0,1}$ of
$H^0(X,\frA)$ as
\begin{equation}
\CA^{0,1}\mapsto g\CA^{0,1}g^{-1}+g\dparb g^{-1}~.
\end{equation}
Thus, we have $\CM\cong H^0(X,\frA)/H^0(X,\frS)$ and this is the
Dolbeault description of holomorphic vector bundles.

Contrary to the connections used in the Dolbeault description, the
\v{C}ech description uses transition function to define vector
bundles. Clearly, such a transition function has to belong to the
first \v{C}ech cocyle set of a suitable sheaf $\frG$. Furthermore,
we want to call two vector bundles equivalent, if there exists an
element $h$ of $C^0(\frU,\frG)$ such that
\begin{equation}
f_{ab}=h^{-1}_a\tilde{f}_{ab}h_b~~~\mbox{on
all}~~\CU_a\cap\CU_b\neq\varnothing~.
\end{equation}
Thus, we observe that holomorphic and smooth vector bundles have
transition functions which are elements of the \v{C}ech cohomology
sets $H^1(\frU,\frH)$ and $H^1(\frU,\frS)$, respectively.

If the patches $\CU_a$ of the covering $\frU$ are Stein manifolds,
one can show that the first \v{C}ech cohomology sets are
independent of the covering and depend only on the manifold $X$,
e.g.\ $H^1(\frU,\frS)=H^1(X,\frS)$. This is well known to be the
case for purely bosonic twistor spaces and since the covering is
obviously unaffected by the extension to an infinitesimal
neighborhood,\footnote{An infinitesimal neighborhood cannot be
covered partially.} we refrain from going into too much detail at
this point.

To connect both descriptions, let us first introduce the subset
$\frX$ of $C^0(\frU,\frS)$ given by a collection of $G$-valued
functions $\psi=\{\psi_a\}$, which fulfill
\begin{equation}
\psi_a\dparb\psi_a^{-1}=\psi_b\dparb\psi_b^{-1}~.
\end{equation}
Due to \eqref{eomdol}-\eqref{eomdol2}, elements of $H^0(X,\frA)$
can be written as $\psi\dparb\psi^{-1}$ with $\psi\in\frX$. Thus,
for every $\CA^{0,1}\in H^0(X,\frA)$ we have corresponding
elements $\psi\in\frX$. Now, one of these $\psi$ can be used to
define the transition functions of a rank $n$ holomorphic vector
bundle $\CE$ over $X$ by the formula
\begin{equation}
f_{ab}=\psi^{-1}_a\psi_b~~~\mbox{on}~~\CU_a\cap\CU_b\neq\varnothing~.
\end{equation}
It is easily checked, that the $f_{ab}$ constructed in this way
are holomorphic. Furthermore, they define holomorphic vector
bundles which are topologically trivial, but not holomorphically
trivial. Thus, they belong to the kernel of a map
$\rho:H^1(X,\frH)\rightarrow H^1(X,\frS)$.

The isomorphy between the moduli spaces of both descriptions is
easily found. We have the short exact sequence
\begin{equation}
0\rightarrow\frH\stackrel{i}{\rightarrow}\frS\stackrel{\delta^0}{\rightarrow}\frA
\stackrel{\delta^1}{\rightarrow}0~,
\end{equation}
where $i$ denotes the embedding of $\frH$ in $\frS$, $\delta^0$ is
the map $\frS\ni\psi\mapsto \psi\dparb\psi^{-1}\in\frA$ and
$\delta^1$ is the map $\frA\ni\CA^{0,1}\mapsto \dparb
\CA^{0,1}+\CA^{0,1}\wedge\CA^{0,1}$ (cf.\ \cite{dolbeault} for the
purely bosonic case). This short exact sequence induces a long
exact sequence of cohomology groups
\begin{equation}
0\rightarrow H^0(X,\frH)\stackrel{i_*}{\rightarrow}
H^0(X,\frS)\stackrel{\delta^0_*}{\rightarrow}
H^0(X,\frA)\stackrel{\delta^1_*}{\rightarrow}
H^1(X,\frH)\stackrel{\rho}{\rightarrow} H^1(X,\frS)\rightarrow
...~,
\end{equation}
and from this we see that $\kernel\,\rho\cong
H^0(X,\frA)/H^0(X,\frS)$. Thus, the moduli spaces of both
descriptions are isomorphic, i.e.\ the moduli space of holomorphic
vector bundles smoothly equivalent to the trivial bundle is
bijective to the moduli space of hCS theory on $X$.

\section{Twistor correspondence for fattened complex manifolds}

\subsection{Twistor geometry}

The purpose of this section is to fix our notation. For a more
extensive review on twistors, supertwistors and the twistor
correspondence in conventions very similar to the ones used here,
see e.g.\ \cite{Popov:2004rb} and the references therein.

Consider the projective space $\CPP^3$ described by homogeneous
coordinates $(\omega^\alpha,\lambda_\ald)$ with indices
$\alpha,\ald=1,2$. Let $\CP^3$ be the subspace of $\CPP^3$ for
which $(\lambda_\ald)\neq(0,0)^T$ and let it be covered by two
patches $U_+$ and $U_-$ for which $\lambda_{\dot{1}}\neq 0$ and
$\lambda_{\dot{2}}\neq 0$, respectively. Then we can introduce
inhomogeneous coordinates
$(z_+^\alpha,\lambda_+):=(\omega^\alpha/\lambda_{\dot{1}},\lambda_{\dot{2}}/\lambda_{\dot{1}})$
on $U_+$ and
$(z_-^\alpha,\lambda_-):=(\omega^\alpha/\lambda_{\dot{2}},\lambda_{\dot{1}}/\lambda_{\dot{2}})$
on $U_-$. Note that $z^\alpha_+=\lambda_+z^\alpha_-$ and
$\lambda_+=(\lambda_-)^{-1}$ on $U_+\cap U_-$, and thus the space
$\CP^3$ is the rank two vector bundle $\CO(1)\oplus\CO(1)$ over
the Riemann sphere $\CPP^1$. Holomorphic sections of this bundle
are parametrized by elements $(x^{\alpha\ald}$) of the space
$\FC^4$ and locally defined by the equations
\begin{equation}
z_+^\alpha=x^{\alpha\ald}\lambda^+_\ald\eand
z_-^\alpha=x^{\alpha\ald}\lambda^-_\ald~,
\end{equation}
where we introduced the notation
$(\lambda^+_\ald)=(1,\lambda_+)^T$ and
$(\lambda^-_\ald)=(\lambda_-,1)^T$. Using these equations, one can
establish a double fibration
\begin{equation}\label{dblfibration}
\begin{picture}(50,40)
\put(0.0,0.0){\makebox(0,0)[c]{$\CP^3$}}
\put(64.0,0.0){\makebox(0,0)[c]{$\FC^4$}}
\put(34.0,33.0){\makebox(0,0)[c]{$\FC^4\times \CPP^1$}}
\put(7.0,18.0){\makebox(0,0)[c]{$\pi_2$}}
\put(55.0,18.0){\makebox(0,0)[c]{$\pi_1$}}
\put(25.0,25.0){\vector(-1,-1){18}}
\put(37.0,25.0){\vector(1,-1){18}}
\end{picture}
\end{equation}
with obvious projections $\pi_1$ and $\pi_2$. The tangent bundle
$T^{(0,1)}\CP^3$ has local sections $\der{\bar{z}^\alpha_\pm}$ and
$\der{\bl_\pm}$ and the (1,0)-part of the tangent space to the
leaves of the projection $\pi_2$, i.e.\ $\kernel\, \pi_{2*}$, is
locally spanned by
$V^\pm_\alpha=\lambda_\pm^\ald\der{x^{\alpha\ald}}$, where we used
$\lambda_\pm^\ald=\eps^{\ald\bed}\lambda^\pm_\bed$ and
$\eps^{\dot{1}{\dot{2}}}=-1$.

To obtain the $\CN$-extended supertwistor space
\cite{Ferber:1977qx}, we replace the rank two vector bundle
$\CP^3=\FC^2\otimes\CO(1)$ by\footnote{The operator $\Pi$ inverts
the parity of the fibre coordinates in a fibre bundle, see
\cite{Cartier:2002zp,Popov:2004rb}.}
$\CP^{3|\CN}:=\FC^2\otimes\CO(1)\oplus\Pi\left(\FC^\CN\otimes\CO(1)\right)$.
This space is again covered by two patches $\CU_+$ and $\CU_-$
with the same bosonic coordinates and sections as $\CP^3$. Local
fermionic coordinates are $\eta_k^\pm$ with
$\eta_k^+=\lambda_+\eta_k^-$ on $\CU_+\cap\CU_-$, where
$k=1,...,\CN$, and fermionic sections are parameterized by
elements $(\eta_k^\ald)$ of the space $\FC^{0|2\CN}$ as
$\eta_k^\pm=\eta_k^\ald\lambda^\pm_\ald$. The total moduli space
of sections of the vector bundle $\CP^{3|\CN}\rightarrow \CPP^1$
is $\FC^{4|2\CN}$ and the double fibration \eqref{dblfibration} is
extended to
\begin{equation}\label{superdblfibration}
\begin{picture}(50,50)
\put(0.0,0.0){\makebox(0,0)[c]{$\CP^{3|\CN}$}}
\put(74.0,0.0){\makebox(0,0)[c]{$\FC^{4|2\CN}$}}
\put(40.0,37.0){\makebox(0,0)[c]{$\FC^{4|2\CN}\times\CPP^1$}}
\put(7.0,20.0){\makebox(0,0)[c]{$\pi_2$}}
\put(65.0,20.0){\makebox(0,0)[c]{$\pi_1$}}
\put(25.0,27.0){\vector(-1,-1){18}}
\put(47.0,27.0){\vector(1,-1){18}}
\end{picture}
\end{equation}
Furthermore, we get the additional fermionic vector fields
$\der{\bar{\eta}_k^\pm}$ on $\CP^{3|\CN}$ and the $(1,0)$-vector
fields $D^k_\pm:=\lambda^\ald_\pm\der{\eta^\ald_k}$ along the
leaves of the projection $\pi_2$.

Instead of the shorthand notation $\CP^{3|\CN}$, we will often
write $(\CP^3,\CO_{[\CN]})$ in the following, which makes the
extension of the structure sheaf of $\CP^3$ explicit. The sheaf
$\CO_{[\CN]}$ is locally the tensor product of the structure sheaf
of a patch of the assigned space $\CP^3$ (which is suppressed in
our notation) and a Gra{\ss}mann algebra of $\CN$ generators.

Let us roughly outline a typical twistor correspondence. For this,
consider again the double fibration \eqref{dblfibration}. We start
with the Dolbeault description of a trivial complex vector bundle
$E$ over $\CP^3$, i.e.\ we consider solutions $A^{0,1}$ to the
equations of motion $\dparb A^{0,1}+A^{0,1}\wedge A^{0,1}=0$ of
hCS theory on $\CP^3$. These solutions can be written as
$A^{0,1}_\pm=\psi_\pm\dparb\psi_\pm^{-1}$ on each patch, and the
functions $\psi_\pm$ can be used to construct a transition
function $f_{+-}=\psi_+^{-1}\psi_-$ for a holomorphic vector
bundle $\tilde{E}$, which is smoothly equivalent to $E$. Thus we
switched from a bundle in the Dolbeault description to an
equivalent bundle in the \v{C}ech description. In the following,
we restrict ourselves to solutions $A^{0,1}$ for which the
component $A^{0,1}_{\bl_\pm}:=\der{\bl_\pm}\lrcorner A^{0,1}$
vanishes. These correspond to bundles $\tilde{E}$, which are
holomorphically trivial when restricted to any projective line
$\CPP^1_x\embd\CP^3$, a demand which is essential in the
subsequent construction. After pulling this bundle back to
$\FC^4\times\CPP^1$, we use the holomorphical triviality of
$\tilde{E}|_{\CPP^1_x}$ to choose a different gauge
$\psi\rightarrow \hat{\psi}$ in which the $\hat{\psi}$s are purely
holomorphic functions of the coordinates on $\FC^4\times\CPP^1$.
From these $\hat{\psi}$, we obtain again a (gauge transformed)
Dolbeault description $\hat{A}^\pm_\alpha=\hat{\psi}_\pm
V_\alpha^\pm\hat{\psi}_\pm^{-1}$ of the bundle $E$ pulled back to
$\FC^4\times \CPP^1$, which can be readily pushed forward to
$\FC^4$. Then the gauge potential obtained in this way satisfies
the self-duality equations on $\FC^4$. Altogether, there is a
one-to-one correspondence between the moduli space of solutions
(with $A^{0,1}_{\bl_\pm}=0$) to the hCS equations of motion on
$\CP^3$ on the one hand side and the moduli space of solutions to
the SDYM equations on $\FC^4$ on the other hand side, where the
moduli spaces are obtained by dividing the solution spaces by the
respective group of gauge transformations.

\subsection{From the supertwistor space $\CP^{3|4}$ to fattened
complex manifolds}

The Calabi-Yau property, i.e.\ vanishing of the first Chern class
or equivalently the existence of a globally well-defined
holomorphic volume form, is essential for defining the B-model on
a certain space. Consider the space $\CP^{3|4}$ as introduced in
the last section. Since the volume element $\Omega$ which is
locally given by $\Omega_\pm:=\pm\dd z_\pm^1\wedge\dd
z^2_\pm\wedge \dd \lambda_\pm\wedge\dd
\eta_1^\pm\wedge...\wedge\dd \eta_4^\pm$ is globally defined and
holomorphic, $\CP^{3|4}$ is a Calabi-Yau
supermanifold.\footnote{The total Chern class of $\CP^{3|4}$
indeed adds up to zero: The body
$\CO(1)\oplus\CO(1)\rightarrow\CPP^1$ contributes 1 from each of
the fibres of the line subbundles and 2 from the cotangent space
of the sphere. The fermionic fibres contribute $-4$ altogether, as
$\Pi\CO(1)$ has Chern class $-1$ due to the inverse appearance of
the equivalent of the Jacobi determinant in Berezin integration.}
Other spaces which have a twistorial $\CO(1)\oplus\CO(1)$ body and
are still Calabi-Yau supermanifolds are, e.g., the weighted
projective spaces\footnote{In fact, one rather considers their
open subspaces $W\CPP^{3|2}(1,1,1,1|p,q)\backslash
W\CPP^{1|2}(1,1|p,q)$.} $W\CPP^{3|2}(1,1,1,1|p,q)$ with $(p,q)$
equal to $(1,3)$, $(2,2)$ and $(4,0)$ as considered in
\cite{Popov:2004nk}. The B-model on these manifolds was shown to
be equivalent to $\CN=4$ SDYM theory with a truncated field
content. Additionally in the cases $(2,2)$ and $(4,0)$, the parity
of some fields is changed, similarly to the result of a
topological twist.

An obvious idea to obtain even more Calabi-Yau supermanifolds
directly from $\CP^{3|4}$ is to combine several fermionic
variables into a single one,\footnote{A similar situation has been
considered in \cite{Lechtenfeld:2004cc}, where all the fermionic
variables where combined into a single even nilpotent one.} e.g.\
to consider coordinates $(\zeta_1:=\eta_1,$
$\zeta_2:=\eta_2\eta_3\eta_4)$. In an analogous situation for
bosonic variables, one could always at least locally find
additional coordinates complementing the reduced set to a set
describing the full space. Fixing the complementing coordinates to
certain values then means, that one considers a subvariety of the
full space. However, as there is no inverse of Gra{\ss}mann variables,
the situation here is different. Instead of taking a subspace, we
rather restrict the algebra of functions (and similarly the set of
differential operators) by demanding a certain dependence on the
Gra{\ss}mann variables. One can indeed find complementing sets of
functions to restore the full algebra of functions on $\CP^{3|4}$.
Underlining the argument that we do not consider a subspace of
$\CP^{3|4}$ is the observation that we still have to integrate
over the full space $\CP^{3|4}$: $\int \dd \zeta_1\dd\zeta_2=\int
\dd \eta_1...\dd \eta_4$. This picture has a slight similarity to
the definition of the body of a supermanifold as given in
\cite{DeWitt:1992cy,Cartier:2002zp}.

Possible inequivalent groupings of the Gra{\ss}mann coordinates of
$\CP^{3|4}$ are the previously given example
$(\zeta_1:=\eta_1,\zeta_2:=\eta_2\eta_3\eta_4)$ as well as
$(\zeta_1=\eta_1,\zeta_2=\eta_2,\zeta_3=\eta_3\eta_4)$,
$(\zeta_1=\eta_1\eta_2,$ $\zeta_2=\eta_3\eta_4)$, and
$(\zeta_1=\eta_1\eta_2\eta_3\eta_4)$. They correspond to exotic
supermanifolds of dimension $(3\oplus 0|2)$, $(3\oplus 1|2)$,
$(3\oplus 2|0)$, and $(3\oplus 1|0)$, respectively. Considering
hCS theory on them, one finds that the first one is equivalent to
the case $W\CPP^{3|2}(1,1,1,1|1,3)$ which was already discussed in
\cite{Popov:2004nk}. The case $(3\oplus 2|0)$ will be similar to a
the case $W\CPP^{3|2}(1,1,1,1|2,2)$, but with a field content of
partially different parity. The case $(3\oplus 1|2)$ is a mixture
easily derived from combining the full case $\CP^{3|4}$ with the
case $(3\oplus 2|0)$. We restrict ourselves in the following to
the cases $(3\oplus 2|0)$ and $(3\oplus 1|0)$.

Instead of considering independent twistor correspondences between
fattened complex manifolds and the moduli space of relative
deformations of the contained $\CPP^1$, we will focus on {\em
reductions} of the correspondence between $\CP^{3|4}$ and
$\FC^{4|8}$. This formulation allows for a more direct
identification of the remaining subsectors of $\CN=4$ self-dual
Yang-Mills theory and can in a sense be understood as a fermionic
dimensional reduction.

\subsection{The B-model on $\SA$}

\subsubsection{Geometrical considerations}

The starting point of our discussion is the supertwistor space
$\CP^{3|4}=(\CP^3,\CO_{[4]})$. Consider the differential operators
\begin{equation}
\CD^{i1}_\pm:=\eta_1^\pm\eta_2^\pm\der{\eta_i^\pm}\eand
\CD^{i2}_\pm:=\eta_3^\pm\eta_4^\pm\der{\eta_{i+2}^\pm}~~~\mbox{for}~~i=1,2~~,
\end{equation}
which are maps $\CO_{[4]}\rightarrow\CO_{[4]}$. The space
$\CP^{3}$ together with the structure sheaf
\begin{equation}
\CO_{(1,2)}:=\bigcap_{i,j=1,2}\ker
\CD^{ij}_+=\bigcap_{i,j=1,2}\ker \CD^{ij}_-~,
\end{equation}
which is a reduction of $\CO_{[4]}$, is the fattened complex
manifold $\SA$, covered by two patches $\CU_+$ and $\CU_-$ and
described by local coordinates
$(z^\alpha_\pm,\,\lambda_\pm,\,y^\pm_1:=\eta^\pm_1\eta^\pm_2,\,
y^\pm_2:=\eta^\pm_3\eta^\pm_4)$. The two even nilpotent
coordinates $y^\pm_i$ are each sections of the line bundle
$\CO(2)$ with the identification $(y^\pm_i)^2\sim 0$.

As pointed out before, the coordinates $y^\pm_i$ do not allow for
a complementing set of coordinates, and therefore it is not
possible to use Leibniz calculus in the transition from the
$\eta$-coordinates on $(\CP^{3},\CO_{[4]})$ to the $y$-coordinates
on $(\CP^{3},\CO_{(1,2)})$. Instead, from the observation that
\begin{align}\nonumber
\eta_2^\pm\der{y^\pm_1}=\left.\der{\eta^\pm_1}\right|_{\CO_{(1,2)}}~,~~~
\eta_1^\pm\der{y^\pm_1}=-\left.\der{\eta^\pm_2}\right|_{\CO_{(1,2)}}~,\\\label{derid}
\eta_4^\pm\der{y^\pm_2}=\left.\der{\eta^\pm_3}\right|_{\CO_{(1,2)}}~,~~~
\eta_3^\pm\der{y^\pm_2}=-\left.\der{\eta^\pm_4}\right|_{\CO_{(1,2)}}~,
\end{align}
one directly obtains the following identities on
$(\CP^{3},\CO_{(1,2)})$:
\begin{equation}\label{deridb}
\der{y^\pm_1}=\der{\eta^\pm_2}\der{\eta^\pm_1}\eand
\der{y^\pm_2}=\der{\eta^\pm_4}\der{\eta^\pm_3}~.
\end{equation}
Equations \eqref{derid} are easily derived by considering an
arbitrary section $f$ of $\CO_{(1,2)}$:
\begin{equation}
f=a^0+a^1 y_1+a^2 y_2+a^{12}
y_1y_2=a^0+a^1\eta_1\eta_2+a^2\eta_3\eta_4+a^{12}\eta_1\eta_2\eta_3\eta_4~,
\end{equation}
where we suppressed the $\pm$ labels for convenience. Acting,
e.g., by $\der{\eta_1}$ on $f$, we see that this equals an action
of $\eta_2\der{y_1}$. It is then also obvious, that we can make
the formal identification \eqref{deridb} on
$(\CP^{3},\CO_{(1,2)})$. Still, a few more comments on
\eqref{deridb} are in order. These differential operators clearly
map $\CO_{(1,2)}\rightarrow \CO_{(1,2)}$ and fulfill
\begin{equation}\label{derdel}
\der{y^\pm_i}\,y^\pm_j=\delta^i_j~.
\end{equation}
Note, however, that they do not quite satisfy the Leibniz rule,
e.g.:
\begin{equation}
1=\der{y^\pm_1}\,y^\pm_1=\der{y^\pm_1}\,(\eta^\pm_1\eta^\pm_2)
\neq
\left(\der{y^\pm_1}\,\eta^\pm_1\right)\eta^\pm_2+\eta^\pm_1\left(\der{y^\pm_1}\,\eta^\pm_2\right)=0~.
\end{equation}
This does not affect the fattened complex manifold $\SA$ at all,
but it imposes an obvious constraint on the formal manipulation of
expressions involving the $y$-coordinates rewritten in terms of
the $\eta$-coordinates.

For the cotangent space, we have the identification $\dd
y^\pm_1=\dd \eta^\pm_2\dd \eta^\pm_1$ and $\dd y^\pm_2=\dd
\eta^\pm_4\dd \eta^\pm_3$ and similarly to above, one has to take
care in formal manipulations, as integration is equivalent to
differentiation.

As discussed in section 3.1, holomorphic sections of the bundle
$\CP^{3|4}\rightarrow \CPP^1$ are described by moduli which are
elements of the space $\FC^{4|8}=(\FC^4,\CO_{[8]})$. After the
above reduction, holomorphic sections of the bundle
$\SA\rightarrow\CPP^1$ are defined by the equations
\begin{equation}\label{secsA}
z^\ald_\pm=x^{\alpha\ald}\lambda_\ald^\pm\eand
y^\pm_i=y^{\ald\bed}_i\lambda^\pm_\ald\lambda^\pm_\bed~.
\end{equation}
While the Gra{\ss}mann algebra of the coordinates $\eta_k^\pm$ of
$\CP^{3|4}$ immediately imposed a Gra{\ss}mann algebra on the moduli
$\eta_k^\ald\in\FC^{0|8}$, the situation here is more subtle. We
have\footnote{The brackets $(\cdot)$ and $[\cdot]$ denote
symmetrization and antisymmetrization, respectively, of the
enclosed indices with appropriate weight.}
$y_1^{(\ald\bed)}=\eta_1^{(\ald}\eta_2^{\bed)}$ and from this, we
already note that $(y_1^{\dot{1}\dot{2}})^2\neq0$ but only
$(y_1^{\dot{1}\dot{2}})^3=0$. Thus, the moduli space is a
fattening of order 1 in $y_1^{\dot{1}\dot{1}}$ and
$y_1^{\dot{2}\dot{2}}$, but a fattening of order 2 in
$y_1^{\dot{1}\dot{2}}$ which analogously holds for
$y_2^{\ald\bed}$. Furthermore, we have the additional identities
\begin{equation}\label{id1}
y_i^{\dot{1}\dot{2}}y_i^{\dot{1}\dot{2}}=-\tfrac{1}{2}
y_i^{\dot{1}\dot{1}}y_i^{\dot{2}\dot{2}}~~~\mbox{and}~~~
y_i^{\dot{1}\dot{2}}y_i^{\dot{2}\dot{2}}=y_i^{\dot{1}\dot{2}}y_i^{\dot{1}\dot{1}}=0~.
\end{equation}
Additional conditions which appear when working with fattened
complex manifolds are not unusual and similar problems were
encountered, e.g., in the discussion of fattened ambitwistor
spaces in \cite{Eastwood:1987}.

More formally, one can introduce the differential operators
\begin{align}
&\CD^{1c}=(\eta_1^\ald\dpar^1_\ald-\eta_2^\ald\dpar^2_\ald)~,~~~
\CD^{2c}=(\eta_3^\ald\dpar^3_\ald-\eta_4^\ald\dpar^4_\ald)~,\\
&\CD^{1s}=(\dpar^2_{\dot{1}}\dpar^1_{\dot{2}}-\dpar^2_{\dot{2}}\dpar^1_{\dot{1}})~,~~~
\CD^{2s}=(\dpar^4_{\dot{1}}\dpar^3_{\dot{2}}-\dpar^4_{\dot{2}}\dpar^3_{\dot{1}})~
\end{align}
which map $\CO_{[8]}\rightarrow\CO_{[8]}$, and consider the
overlap of kernels
\begin{equation}
\CO_{(1;2,6)}:=\bigcap_{i=1,2}\left(\ker(\CD^{ic})\cap\ker(\CD^{is})\right)~.
\end{equation}
The space $\FC^{4}$ together with the structure sheaf
$\CO_{(1;2,6)}$, which is a reduction of $\CO_{[8]}$, is exactly
the moduli space described above, i.e.\ a fattened complex
manifold $\FC^{4\oplus 6|0}$ on which the coordinates
$y_i^{\ald\bed}$ satisfy the additional constrains \eqref{id1}.

Altogether, we have the following reduction of the full double
fibration \eqref{superdblfibration} for $\CN=4$:
\begin{equation}\label{dblfibrationA}
\begin{picture}(50,45)
\put(0.0,0.0){\makebox(0,0)[c]{$(\CP^{3},\CO_{[4]})$}}
\put(64.0,0.0){\makebox(0,0)[c]{$(\FC^{4},\CO_{[8]})$}}
\put(34.0,39.0){\makebox(0,0)[c]{$(\FC^{4}\times\CPP^1,\CO_{[8]}\otimes\CO_{\CPP^1})$}}
\put(7.0,21.0){\makebox(0,0)[c]{$\pi_2$}}
\put(56.0,21.0){\makebox(0,0)[c]{$\pi_1$}}
\put(25.0,28.0){\vector(-1,-1){18}}
\put(37.0,28.0){\vector(1,-1){18}}
\end{picture}\hspace{2.3cm}\longrightarrow\hspace{2.1cm}
\begin{picture}(50,45)
\put(0.0,0.0){\makebox(0,0)[c]{$(\CP^{3},\CO_{(1,2)})$}}
\put(79.0,0.0){\makebox(0,0)[c]{$(\FC^{4},\CO_{(1;2,6)})$}}
\put(42.0,39.0){\makebox(0,0)[c]{$(\FC^{4}\times\CPP^1,\CO_{(1;2,6)}\otimes\CO_{\CPP^1})$}}
\put(7.0,21.0){\makebox(0,0)[c]{$\pi_2$}}
\put(56.0,21.0){\makebox(0,0)[c]{$\pi_1$}}
\put(25.0,28.0){\vector(-1,-1){18}}
\put(37.0,28.0){\vector(1,-1){18}}
\end{picture}
\end{equation}
where $\CO_{\CPP^1}$ is the structure sheaf of the Riemann sphere
$\CPP^1$. The tangent spaces along the leaves of the projection
$\pi_2$ are spanned by the vector fields
\begin{align}\nonumber
&V_\alpha^\pm=\lambda_\pm^\ald\dpar_{\alpha\ald}~,
&&V_\alpha^\pm=\lambda_\pm^\ald\dpar_{\alpha\ald}~,\\
&D^k_\pm=\lambda^\ald_\pm\der{\eta_k^\ald}~, &&D^i_{\bed
\pm}=\lambda^\ald_\pm\der{y_i^{(\ald\bed)}}
\end{align}
in the left and right double fibration in \eqref{dblfibrationA},
where $k=1,...,4$. Note that similarly to \eqref{derid}, we have
the identities
\begin{align}\nonumber
\eta_2^\ald\der{y_1^{(\ald\bed)}}=\left.\der{\eta_1^\bed}\right|_{\CO_{(1;2,6)}}~,~~~
\eta_1^\ald\der{y_1^{(\ald\bed)}}=-\left.\der{\eta_2^\bed}\right|_{\CO_{(1;2,6)}}~,\\\label{derid2}
\eta_4^\ald\der{y_2^{(\ald\bed)}}=\left.\der{\eta_3^\bed}\right|_{\CO_{(1;2,6)}}~,~~~
\eta_3^\ald\der{y_2^{(\ald\bed)}}=-\left.\der{\eta_4^\bed}\right|_{\CO_{(1;2,6)}}~,
\end{align}
and it follows, e.g., that
\begin{equation}
\left.D^1_\pm\right|_{\CO_{(1;2,6)}}=\eta_2^\ald
D^1_{\ald\pm}\eand
\left.D^2_\pm\right|_{\CO_{(1;2,6)}}=-\eta_1^\ald D^1_{\ald\pm}~.
\end{equation}

\subsubsection{Field theoretical considerations}

The topological B-model on $\SA=(\CP^{3},\CO_{(1,2)})$ is
equivalent to hCS theory on $\SA$ since a reduction of the
structure sheaf does not affect the arguments used for this
equivalence in \cite{Witten:1992fb,Witten:2003nn}. Consider a
trivial rank $n$ complex vector bundle\footnote{Note that the
components of sections of ordinary vector bundles over a
supermanifold are superfunctions. The same holds for the
components of connections and transition functions.} $\CE$ over
$\SA$ with a connection $\CA$. The action for hCS theory on this
space reads
\begin{equation}
S=\int_{\SA_{\mathrm{ch}}} \Omega^{3\oplus
2|0}\wedge\tr\left(\CA^{0,1}\wedge\bar{\dpar}\CA^{0,1}+
\tfrac{2}{3}\CA^{0,1}\wedge\CA^{0,1}\wedge\CA^{0,1}\right)~,
\end{equation}
where $\SA_{\mathrm{ch}}$ is the subspace\footnote{This
restriction to a subspace holomorphic in the fermionic
coordinates, i.e.\ a chiral subspace, was proposed in
\cite{Witten:2003nn} and is related to self-duality.} of $\SA$ for
which $\bar{y}^\pm_i=0$, $\CA^{0,1}$ is the (0,1)-part of $\CA$
and $\Omega^{3\oplus 2|0}$ is the holomorphic volume form, e.g.\
$\Omega^{3\oplus 2|0}_+=\dd z^1_+\wedge\dd z^2_+\wedge\dd
\lambda_+\wedge \dd y_1^+\wedge \dd y_2^+$. The equations of
motion read $\dparb \CA^{0,1}+\CA^{0,1}\wedge \CA^{0,1}=0$ and
solutions define a holomorphic structure $\dparb_\CA$ on $\CE$.
Given such a solution $\CA^{0,1}$, one can locally write
$\CA^{0,1}|_{\CU_\pm}=\psi_\pm\dparb\psi^{-1}_\pm$ with regular
matrix-valued functions $\psi_\pm$ smooth on the patches $\CU_\pm$
and from the gluing condition
$\psi_+\dparb\psi^{-1}_+=\psi_-\dparb\psi^{-1}_-$ on the overlap
$\CU_+\cap\CU_-$, one obtains
$\dparb\left(\psi^{-1}_+\psi_-\right)=0$. Thus,
$f_{+-}:=\psi^{-1}_+\psi_-$ defines a transition function for a
holomorphic vector bundle $\tilde{\CE}$, which is (smoothly)
equivalent to $\CE$.

Consider now the pull-back of the bundle $\tilde{\CE}$ along
$\pi_2$ in \eqref{dblfibrationA} to the space $\FC^{4|8}\times
\CPP^1$, i.e.\ the holomorphic vector bundle $\pi_2^*\tilde{\CE}$
with transition function $\pi_2^* f_{+-}$ satisfying $V^\pm_\alpha
\left(\pi_2^*f_{+-}\right)=D^k_{\pm} \left(\pi_2^*f_{+-}\right)=0$.
Let us suppose that the vector bundle $\pi_2^*\tilde{\CE}$ becomes
holomorphically trivial\footnote{This assumption is crucial for
the Penrose-Ward transform and reduces the space of possible
$\CA^{0,1}$ to an open subspace around $\CA^{0,1}=0$.} when
restricted to sections $\CPP^1_{x,y}\embd\CP^{3|4}$. This implies,
that there is a splitting $\pi^*_2
f_{+-}=\hat{\psi}_+^{-1}\hat{\psi}_-$, where $\hat{\psi}_\pm$ are
group-valued functions which are holomorphic in the moduli
$(x^{\alpha\ald},\eta^{\ald}_k)$ and $\lambda_\pm$. From the
condition $V^\pm_\alpha \left(\pi_2^*f_{+-}\right)=D_\pm^{k}
\left(\pi_2^*f_{+-}\right)=0$ we obtain, e.g.\ on $\CU_+$
\begin{align}\nonumber
\hat{\psi}_+
V^+_\alpha\hat{\psi}_+^{-1}=\hat{\psi}_-V^+_\alpha\hat{\psi}_-^{-1}=:\lambda^\ald_+
\hat{\CA}_{\alpha\ald}&=:\hat{\CA}^+_\alpha~,\\\nonumber
\hat{\psi}_+ D^{k}_+
\hat{\psi}_+^{-1}=\hat{\psi}_-D^{k}_+\hat{\psi}_-^{-1}=:\lambda^\ald_+
\hat{\CA}^k_{\ald}&=:\hat{\CA}^{k}_{+}~,\\\nonumber \hat{\psi}_+
\dpar_{\bl_+}
\hat{\psi}_+^{-1}=\hat{\psi}_-\dpar_{\bl_+}\hat{\psi}_-^{-1}&=:\hat{\CA}_{\bl_+}=0~,\\\label{defA}
\hat{\psi}_+
\dpar_{\bar{x}^{\alpha\ald}}\hat{\psi}_+^{-1}=\hat{\psi}_-\dpar_{\bar{x}^{\alpha\ald}}\hat{\psi}_-^{-1}
&=0~.
\end{align}
Considering the reduced structure sheaves, we can rewrite the
second line of \eqref{defA}, e.g.\ for $k=1$ as
\begin{equation}\label{defA2}
\eta_2^\bed\hat{\psi}_+ D^{1}_{\bed+}
\hat{\psi}_+^{-1}=\eta_2^\bed\hat{\psi}_-D^{1}_{\bed+}\hat{\psi}_-^{-1}=:
\eta_2^\bed\lambda^\ald_+ \hat{\CA}^1_{\ald\bed}~,
\end{equation}
which yields $\eta_2^\bed
\hat{\CA}^1_{\ald\bed}=\hat{\CA}^1_{\ald}$. From this equation
(and similar ones for other values of $k$) and the well-known
superfield expansion of $\hat{\CA}^k_{\ald}$ (see e.g.\
\cite{Devchand:1996gv}), one can now construct the superfield
expansion of $\hat{\CA}^i_{\ald\bed}$ by dropping all the terms,
which are not in the kernel of the differential operators
$\CD^{jc}$ and $\CD^{js}$. This will give rise to a bosonic
subsector of $\CN=4$ SDYM theory.

To be more explicit, we can also use \eqref{defA2} and introduce
the covariant derivative
$\nabla_{\alpha\ald}:=\dpar_{\alpha\ald}+[\hat{\CA}_{\alpha\ald},\cdot]$
and the first order differential operator
$\nabla_{\ald\bed}^i:=\dpar^i_{\ald\bed}+[\hat{\CA}^i_{\ald\bed},\cdot]$,
which allow us to rewrite the compatibility conditions of the
linear system behind \eqref{defA}, \eqref{defA2} for the reduced
structure sheaf as
\begin{align}\nonumber
[\nabla_{\alpha\ald},\nabla_{\beta\bed}]+[\nabla_{\alpha\bed},
\nabla_{\beta\ald}]&=0~,&
\eta^\gad_m\left([\nabla_{\ald\gad}^i,\nabla_{\beta\bed}]+[\nabla_{\bed\gad}^i,\nabla_{\beta\ald}]\right)=0~,
\\\eta^\gad_m\eta^\ded_n\left([\nabla_{\ald\gad}^i,\nabla_{\bed\ded}^j]+
[\nabla_{\bed\gad}^i,\nabla_{\ald\ded}^j]\right)&=0~,
\label{compconA}
\end{align}
where $m=2i-1,2i$ and $n=2j-1,2j$. Note that $\nabla_{\ald\bed}^i$
is no true covariant derivative, as $\dpar_{\ald\bed}^i$ and
$\hat{\CA}^i_{\ald\bed}$ do not have the same symmetry properties
in the indices. Nevertheless, the differential operators
$\nabla_{\alpha\ald}$ and $\nabla_{\ald\bed}^i$ satisfy the
Bianchi identities on $(\FC^{4},\CO_{(1;2,6)})$.

By eliminating all $\lambda$-dependence, we have implicitly
performed the push-forward of $\hat{\CA}$ along $\pi_1$ onto
$(\FC^{4},\CO_{(1;2,6)})$. Let us define further tensor
superfields, which could roughly be seen as extensions of the
supercurvature fields and which capture the solutions to the above
equations:
\begin{align}\nonumber
&[\nabla_{\alpha\ald},\nabla_{\beta\bed}]=:\eps_{\ald\bed}
\CF_{\alpha\beta}~,~~~
[\nabla^i_{\ald\gad},\nabla_{\beta\bed}]=:\eps_{\ald\bed}
\CF^i_{\beta\gad}~,\\
&[\nabla^i_{\ald\gad},\nabla^j_{\bed\ded}]=:\eps_{\ald\bed}
\CF^{ij}_{\gad\ded}~,\label{supercurvature}
\end{align}
where $\CF_{\alpha\beta}=\CF_{(\alpha\beta)}$ and
$\CF^{ij}_{\gad\ded}=\CF^{(ij)}_{(\gad\ded)}+\CF^{[ij]}_{[\gad\ded]}$.
Note, however, that we introduced too many of these components.
Considering the third equation in \eqref{compconA}, one notes that
for $i=j$, the terms symmetric in $\gad,\ded$ vanish trivially.
This means, that the components $\CF^{ii}_{(\gad\ded)}$ are in
fact superfluous and we can ignore them in the following
discussion. The second and third equations in
\eqref{supercurvature} can be contracted with $\eps^{\ald\gad}$
and $\eps^{\bed\ded}$, respectively, which yields
\begin{equation}\label{cordef}
-2\nabla_{\beta\bed}
\hat{\CA}^i_{[\dot{1}\dot{2}]}=\CF^i_{\beta\bed}\eand
-2\nabla_{\ald\gad}^i\hat{\CA}^j_{[\dot{1}\dot{2}]}=\CF^{ij}_{\gad\ald}~.
\end{equation}
Furthermore, using Bianchi identities, one obtains immediately the
following equations:
\begin{equation}\label{eom}
\nabla^{\alpha\bed}\CF^i_{\alpha\gad}=0\eand
\nabla_{\alpha\ald}\CF^{ij}_{\bed\gad}=\nabla^{i}_{\ald\bed}\CF_{\alpha\gad}^{j}~.
\end{equation}
Due to self-duality, the first equation is in fact equivalent to
$\nabla^{\alpha\bed}\nabla_{\alpha\bed}\hat{\CA}^i_{[\dot{1}\dot{2}]}=0$,
as is easily seen by performing all the spinor index sums. From
the second equation, one obtains the field equation
$\nabla_{\alpha}{}^\bed\CF^{(12)}_{\bed\gad}=-
2[\hat{\CA}^{(1}_{[\dot{1}\dot{2}]},\nabla_{\alpha\gad}\hat{\CA}^{2)}_{[\dot{1}\dot{2}]}]$
after contracting with $\eps^{\ald\bed}$.

To analyze the actual field content of this theory, we choose {\em
transverse gauge} as in \cite{Harnad:1984vk}, i.e.\ we demand
\begin{equation}
\eta_k^\ald\hat{\CA}^k_\ald=0~.
\end{equation}
This choice reduces the group of gauge transformations to
ordinary, group-valued functions on the body of $\FC^{4|8}$. By
using the identities $\eta_2^\bed
\hat{\CA}^1_{\ald\bed}=\hat{\CA}^1_{\ald}$ etc., one sees that the
above transverse gauge is equivalent to the transverse gauge for
the reduced structure sheaf:
\begin{equation}
y_i^{\ald\bed}\hat{\CA}_{\ald\bed}^i=\eta^{(\ald}_1\eta^{\bed)}_2\hat{\CA}^1_{\ald\bed}+
\eta^{(\ald}_3\eta^{\bed)}_4\hat{\CA}^2_{\ald\bed}=0~.
\end{equation}

In the expansion in the $y$s, the lowest components of
$\CF_{\alpha\beta}$, $\hat{\CA}^i_{[\dot{1}\dot{2}]}$ and
$\CF^{(12)}_{(\ald\bed)}$ are the self-dual field strength
$f_{\alpha\beta}$, two complex scalars $\phi^i$ and the auxiliary
field $G_{\ald\bed}$, respectively. The two scalars $\phi^i$ can
be seen as remainders of the six scalars contained in the $\CN=4$
SDYM multiplet, which will become even clearer in the real case.
The remaining components $\hat{\CA}^i_{(\ald\bed)}$ vanish to
zeroth order in the $y$s due to the choice of transverse gauge.
The field $\CF^i_{\alpha\ald}$ does not contain any new physical
degrees of freedom, as seen from the first equation in
\eqref{cordef}, but it is a composite field. The same holds for
$\CF^{[12]}_{[\gad\ald]}$ as easily seen by contracting the second
equation in \eqref{cordef} by $\eps^{\gad\ald}$:
$\CF^{[12]}_{[\dot{1}\dot{2}]}=-2[\hat{\CA}^1_{[\dot{1}\dot{2}]},\hat{\CA}^2_{[\dot{1}\dot{2}]}]$.

The superfield equations of motion \eqref{eom} are in fact
equivalent to the equations
\begin{equation}\label{eomr}
f_{\ald\bed}=0~,~~~\square
\phi^i=0~,~~~\eps^{\gad\ald}\nabla_{\alpha\ald}G_{\gad\ded}+
2[\phi^{(1},\nabla_{\alpha\ded}\phi^{2)}]=0~.
\end{equation}
To lowest order in the $y$s, the equations obviously match. Higher
orders in the $y$s can be verified by defining the Euler operator
$D:=y_i^{\ald\bed}\nabla_{(\ald\bed)}^i=y_i^{\ald\bed}\dpar_{(\ald\bed)}^i$
and applying $D$ on the superfields and equations of motion which
then turn out to be satisfied, if the equations \eqref{eomr} are
fulfilled.

\subsection{The B-model on $\SB$}

\subsubsection{Geometrical considerations}

The discussion for $\SB$ follows the same lines as for $\SA$ and
is even simpler. Consider again the supertwistor space
$\CP^{3|4}=(\CP^3,\CO_{[4]})$. This time, let us introduce the
following differential operators:
\begin{equation}
\tilde{\CD}_\pm^{kl}:=\eta_k^\pm\der{\eta_l^\pm}~~~\mbox{for}~~k,l=1,...,4~,
\end{equation}
which are maps $\CO_{[4]}\rightarrow\CO_{[4]}$. The space
$\CP^{3}$ together with the extended structure sheaf\footnote{The
same reduction can be obtained by imposing integral constraints
\cite{Lechtenfeld:2004cc}.}
\begin{equation}
\CO_{(1,1)}:=\bigcap_{k\neq l}\ker
\tilde{\CD}^{kl}_+=\bigcap_{k\neq l}\ker \tilde{\CD}^{kl}_-~,
\end{equation}
which is a reduction of $\CO_{[4]}$, is an order one thickening of
$\CP^3$, which we denote by $\SB$. This manifold can be covered by
two patches $\CU_+$ and $\CU_-$ on which we define the coordinates
$(z^\alpha_\pm,\,\lambda_\pm,\,y^\pm:=\eta^\pm_1\eta^\pm_2\eta^\pm_3\eta^\pm_4)$.
The even nilpotent coordinate $y^\pm$ is a section of the line
bundle $\CO(4)$ with the identification $(y^\pm)^2\sim 0$.

Similarly to the case $\SA$, we have the following identities:
\begin{align}\nonumber
&\eta^\pm_2\eta_3^\pm\eta_4^\pm\der{y^\pm}=\left.\der{\eta_1^\pm}\right|_{\CO_{(1,1)}}~,~~~
&\eta^\pm_1\eta_3^\pm\eta_4^\pm\der{y^\pm}=-\left.\der{\eta_2^\pm}\right|_{\CO_{(1,1)}}~,\\
&\eta^\pm_1\eta_2^\pm\eta_4^\pm\der{y^\pm}=\left.\der{\eta_3^\pm}\right|_{\CO_{(1,1)}}~,~~~
&\eta^\pm_1\eta_2^\pm\eta_3^\pm\der{y^\pm}=-\left.\der{\eta_4^\pm}\right|_{\CO_{(1,1)}}~
\end{align}
which lead to the formal identifications
\begin{equation}
\der{y^\pm}=\der{\eta^\pm_4}\der{\eta^\pm_3}\der{\eta^\pm_2}\der{\eta^\pm_1}\eand
\dd
y^\pm=\dd{\eta^\pm_4}\dd{\eta^\pm_3}\dd{\eta^\pm_2}\dd{\eta^\pm_1}~,
\end{equation}
but again with a restriction of the Leibniz rule in formal
manipulations of expressions written in the $\eta$-coordinates as
discussed in the previous section.

The holomorphic sections of the bundle $\SB\rightarrow \CPP^1$ are
defined by the equations
\begin{equation}\label{secsB}
z^\ald_\pm=x^{\alpha\ald}\lambda_\ald^\pm\eand
y^\pm=y^{(\ald\bed\gad\ded)}\lambda^\pm_\ald\lambda^\pm_\bed\lambda^\pm_\gad\lambda^\pm_\ded~.
\end{equation}
From the obvious identification
$y^{(\ald\bed\gad\ded)}=\eta_1^{(\ald}\eta_2^\bed\eta_3^\gad\eta_4^{\ded)}$
we see, that the product
$y^{(\ald\bed\gad\ded)}y^{(\dot{\mu}\dot{\nu}\dot{\rho}\dot{\sigma})}$
will vanish, unless the number of indices equal to $\dot{1}$ is
the same as the number of indices equal to $\dot{2}$. In this
case, we have additionally the identity
\begin{equation}
\sum_{p}(-1)^{n_p} y^{p_1}y^{p_2}=0~,
\end{equation}
where $p$ is a permutation of
$\dot{1}\dot{1}\dot{1}\dot{1}\dot{2}\dot{2}\dot{2}\dot{2}$, $p_1$
and $p_2$ are the first and second four indices of $p$,
respectively, and $n_p$ is the number of exchanges of a $\dot{1}$
and a $\dot{2}$ between $p_1$ and $p_2$, e.g.\
$n_{\dot{1}\dot{1}\dot{1}\dot{2}\dot{1}\dot{2}\dot{2}\dot{2}}=1$.

The more formal treatment is much simpler. We introduce the
differential operators
\begin{align}
&\tilde{\CD}^{klc}=\left(\eta^\ald_l\dpar_\ald^l-\eta^\ald_k\dpar^k_\ald\right)~~~\mbox{without
summation over $k$ and $l$}~,\\
&\tilde{\CD}^{kls}=\left(\dpar_{\dot{1}}^k\dpar_{\dot{2}}^l-\dpar_{\dot{2}}^k\dpar_{\dot{1}}^l\right)~,
\end{align}
which map $\CO_{[8]}\rightarrow\CO_{[8]}$. Then the space
$\FC^{4}$ with the extended structure sheaf $\CO_{(1;2,5)}$
obtained by reducing $\CO_{[8]}$ to the overlap of kernels
\begin{equation}
\CO_{(1;2,5)}:=\bigcap_{k\neq l}
\left(\ker\tilde{\CD}^{klc}\cap\ker\tilde{\CD}^{kls}\right)
\end{equation}
is the moduli space described above. Thus, we have the following
reduction of the full double fibration \eqref{superdblfibration}
for $\CN=4$:
\begin{equation}\label{dblfibrationB}
\begin{picture}(50,45)
\put(0.0,0.0){\makebox(0,0)[c]{$(\CP^{3},\CO_{[4]})$}}
\put(64.0,0.0){\makebox(0,0)[c]{$(\FC^{4},\CO_{[8]})$}}
\put(34.0,39.0){\makebox(0,0)[c]{$(\FC^{4}\times\CPP^1,\CO_{[8]}\otimes\CO_{\CPP^1})$}}
\put(7.0,21.0){\makebox(0,0)[c]{$\pi_2$}}
\put(56.0,21.0){\makebox(0,0)[c]{$\pi_1$}}
\put(25.0,28.0){\vector(-1,-1){18}}
\put(37.0,28.0){\vector(1,-1){18}}
\end{picture}\hspace{2.3cm}\longrightarrow\hspace{2.1cm}
\begin{picture}(50,45)
\put(0.0,0.0){\makebox(0,0)[c]{$(\CP^{3},\CO_{(1,1)})$}}
\put(79.0,0.0){\makebox(0,0)[c]{$(\FC^{4},\CO_{(1;2,5)})$}}
\put(42.0,39.0){\makebox(0,0)[c]{$(\FC^{4}\times\CPP^1,\CO_{(1;2,5)}\otimes\CO_{\CPP^1})$}}
\put(7.0,21.0){\makebox(0,0)[c]{$\pi_2$}}
\put(56.0,21.0){\makebox(0,0)[c]{$\pi_1$}}
\put(25.0,28.0){\vector(-1,-1){18}}
\put(37.0,28.0){\vector(1,-1){18}}
\end{picture}
\end{equation}
where $\CO_{\CPP^1}$ is again the structure sheaf of the Riemann
sphere $\CPP^1$. The tangent spaces along the leaves of the
projection $\pi_2$ are spanned by the vector fields
\begin{align}\nonumber
&V_\alpha^\pm=\lambda_\pm^\ald\dpar_{\alpha\ald}~,
&&V_\alpha^\pm=\lambda_\pm^\ald\dpar_{\alpha\ald}~,\\
&D^k_\pm=\lambda^\ald_\pm\der{\eta_k^\ald}~,
&&D^\pm_{\bed\gad\ded}=\lambda^{\ald}_\pm\dpar_{(\ald\bed\gad\ded)}
\end{align}
in the left and right double fibration in \eqref{dblfibrationB},
where $k=1,...,4$. The further identities
\begin{align}\nonumber
&\eta_2^\bed\eta_3^\gad\eta_4^\ded\der{y^{(\ald\bed\gad\ded)}}=\left.\der{\eta_1^\ald}\right|_{\CO_{(1;2,5)}}~,~~~
&\eta_1^\bed\eta_3^\gad\eta_4^\ded\der{y^{(\ald\bed\gad\ded)}}=-\left.\der{\eta_2^\ald}\right|_{\CO_{(1;2,5)}}~,\\
&\eta_1^\bed\eta_2^\gad\eta_4^\ded\der{y^{(\ald\bed\gad\ded)}}=\left.\der{\eta_3^\ald}\right|_{\CO_{(1;2,5)}}~,~~~
&\eta_1^\bed\eta_2^\gad\eta_3^\ded\der{y^{(\ald\bed\gad\ded)}}=-\left.\der{\eta_4^\ald}\right|_{\CO_{(1;2,5)}}~
\end{align}
are easily derived and from them it follows that e.g.\
\begin{equation}
\left.D^1_\pm\right|_{\CO_{(1;2,5)}}=\eta_2^\bed\eta_3^\gad\eta_4^\ded
D^\pm_{\bed\gad\ded}\eand
\left.D^2_\pm\right|_{\CO_{(1;2,5)}}=-\eta_1^\bed\eta_3^\gad\eta_4^\ded
D^\pm_{\bed\gad\ded}~.
\end{equation}

\subsubsection{Field theoretical considerations}

The topological B-model on $\SB$ is equivalent to hCS theory on
$\SB$ and introducing a trivial rank $n$ complex vector bundle
$\CE$ over $\SB$ with a connection $\CA$, the action reads
\begin{equation}
S=\int_{\SB_{\mathrm{ch}}} \Omega^{3\oplus
1|0}\wedge\tr\left(\CA^{0,1}\wedge\bar{\dpar}\CA^{0,1}+
\tfrac{2}{3}\CA^{0,1}\wedge\CA^{0,1}\wedge\CA^{0,1}\right)~,
\end{equation}
with $\SB_{\mathrm{ch}}$ being the chiral subspace for which
$\bar{y}^\pm=0$ and $\CA^{0,1}$ the $(0,1)$-part of $\CA$. The
holomorphic volume form $\Omega^{3\oplus1|0}$ can be defined,
e.g.\ on $\CU_+$, as $\Omega^{3\oplus1|0}_+=\dd z^1_+\wedge\dd
z^2_+\wedge\dd \lambda_+\wedge \dd y^+$. Following exactly the
same steps as in the case $\SA$, we again obtain the equations
\begin{align}\nonumber
\hat{\psi}_+
V^+_\alpha\hat{\psi}_+^{-1}=\hat{\psi}_-V^+_\alpha\hat{\psi}_-^{-1}=:\lambda^\ald_+
\hat{\CA}_{\alpha\ald}&=:\hat{\CA}^+_\alpha~,\\\nonumber
\hat{\psi}_+ D^{k}_+
\hat{\psi}_+^{-1}=\hat{\psi}_-D^{k}_+\hat{\psi}_-^{-1}=:\lambda^\ald_+
\hat{\CA}^k_{\ald}&=:\hat{\CA}^{k}_{+}~,\\\nonumber \hat{\psi}_+
\dpar_{\bl_+}
\hat{\psi}_+^{-1}=\hat{\psi}_-\dpar_{\bl_+}\hat{\psi}_-^{-1}&=:\hat{\CA}_{\bl_+}=0~,\\\label{defB}
\hat{\psi}_+
\dpar_{\bar{x}^{\alpha\ald}}\hat{\psi}_+^{-1}=\hat{\psi}_-\dpar_{\bar{x}^{\alpha\ald}}\hat{\psi}_-^{-1}
&=0~.
\end{align}
and by considering the reduced structure sheaves, we can rewrite
the second line this time as
\begin{equation}\label{defB2}
\eta_2^\bed\eta_3^\gad\eta_4^\ded\,\hat{\psi}_+
D^+_{\bed\gad\ded}\hat{\psi}_+^{-1}=
\eta_2^\bed\eta_3^\gad\eta_4^\ded\,\hat{\psi}_-D^+_{\bed\gad\ded}\hat{\psi}_-^{-1}
=:\eta_2^\bed\eta_3^\gad\eta_4^\ded\,\lambda^\ald_+
\hat{\CA}_{\ald\bed\gad\ded}=:\eta_2^\bed\eta_3^\gad\eta_4^\ded\,\hat{\CA}^+_{\bed\gad\ded}
\end{equation}
for $k=1$ which yields
$\eta_2^\bed\eta_3^\gad\eta_4^\ded\hat{\CA}_{\ald\bed\gad\ded}=\hat{\CA}^1_{\ald}$.
Similar formul\ae{} are obtained for the other values of $k$, with
which one can determine the superfield expansion of
$\hat{\CA}_{\ald\bed\gad\ded}$ again from the superfield expansion
of $\hat{\CA}^k_\ald$ by dropping the terms which are not in the
kernel of the differential operators $\tilde{\CD}^{klc}$ and
$\tilde{\CD}^{kls}$ for $k\neq l$.

Analogously to the case $\SA$, one can rewrite the linear system
behind \eqref{defB}, \eqref{defB2} for the reduced structure
sheaf. For this, we define the covariant derivative
$\nabla_{\alpha\ald}:=\dpar_{\alpha\ald}+[\hat{\CA}_{\alpha\ald},\cdot]$
and the first order differential operator
$\nabla_{\ald\bed\gad\ded}:=\dpar_{\ald\bed\gad\ded}+[\hat{\CA}_{\ald\bed\gad\ded},\cdot]$.
Then we have
\begin{align}\nonumber
[\nabla_{\alpha\ald},\nabla_{\beta\bed}]+[\nabla_{\alpha\bed},
\nabla_{\beta\ald}]&=0~,\\\nonumber
\eta^{\dot{\nu}}_k\eta^{\dot{\rho}}_m\eta^{\dot{\sigma}}_n\left(
[\nabla_{\dot{\mu}\dot{\nu}\dot{\rho}\dot{\sigma}},\nabla_{\alpha\ald}]+
[\nabla_{\ald\dot{\nu}\dot{\rho}\dot{\sigma}},\nabla_{\alpha\dot{\mu}}]\right)&=0~,
\\
\eta^\bed_r\eta^\gad_s\eta^\ded_t\eta^{\dot{\nu}}_k\eta^{\dot{\rho}}_m\eta^{\dot{\sigma}}_n
\left([\nabla_{\ald\bed\gad\ded},\nabla_{\dot{\mu}\dot{\nu}\dot{\rho}\dot{\sigma}}]+
[\nabla_{\dot{\mu}\bed\gad\ded},\nabla_{\ald\dot{\nu}\dot{\rho}\dot{\sigma}}]\right)
&=0~, \label{compconB}
\end{align}
where $(rst)$ and $(kmn)$ are each a triple of pairwise different
integers between 1 and 4. Again, in these equations the
push-forward $\pi_{1*}\hat{\CA}$ is already implied and solutions
to \eqref{compconB} are captured by the following extensions of
the supercurvature fields:
\begin{align}\nonumber
[\nabla_{\alpha\ald},\nabla_{\beta\bed}]&=:\eps_{\ald\bed}
\CF_{\alpha\beta}~,\\\nonumber
[\nabla_{\dot{\mu}\dot{\nu}\dot{\rho}\dot{\sigma}},\nabla_{\alpha\ald}]&=:\eps_{\ald\dot{\mu}}
\CF_{\alpha\dot{\nu}\dot{\rho}\dot{\sigma}}~,\\
[\nabla_{\ald\bed\gad\ded},\nabla_{\dot{\mu}\dot{\nu}\dot{\rho}\dot{\sigma}}]&=:\eps_{\ald\dot{\mu}}
\CF_{\bed\gad\ded\dot{\nu}\dot{\rho}\dot{\sigma}}~,\label{supercurvatureB}
\end{align}
where $\CF_{\alpha\beta}=\CF_{(\alpha\beta)}$,
$\CF_{\alpha\dot{\nu}\dot{\rho}\dot{\sigma}}=\CF_{\alpha(\dot{\nu}\dot{\rho}\dot{\sigma})}$
and
$\CF_{\bed\gad\ded\dot{\nu}\dot{\rho}\dot{\sigma}}=\CF_{(\bed\gad\ded)(\dot{\nu}\dot{\rho}\dot{\sigma})}$
is symmetric under exchange of
$(\bed\gad\ded)\leftrightarrow(\dot{\nu}\dot{\rho}\dot{\sigma})$.
Consider now the third equation of \eqref{compconB}. Note that the
triples $(rst)$ and $(kmn)$ will have two numbers in common, while
exactly one is different. Without loss of generality, let $r\neq
k$, $s=m$ and $t=n$. Then one easily sees, that the terms
symmetric in $\bed$, $\dot{\nu}$ vanish trivially. This means,
that the field components
$\CF_{\bed\gad\ded\dot{\nu}\dot{\rho}\dot{\sigma}}$ which are
symmetric in $\bed$, $\dot{\nu}$ are again unconstrained
additional fields, which do not represent any of the fields in the
$\CN=4$ SDYM multiplet and we put them to zero, analogously to
$\CF^{ii}_{(\gad\ded)}$ in the case $\SA$.

The second equation in \eqref{supercurvatureB} can be contracted
with $\eps^{\dot{\mu}\dot{\nu}}$ which yields
$2\nabla_{\alpha\ald}\hat{\CA}_{[\dot{1}\dot{2}]\dot{\rho}\dot{\sigma}}=\CF_{\alpha\ald\dot{\rho}\dot{\sigma}}$
and further contracting this equation with $\eps^{\ald\dot{\rho}}$
we have
$\nabla_{\alpha}{}^\ald\hat{\CA}_{[\dot{1}\dot{2}]\ald\dot{\sigma}}=0$.
After contracting the third equation with
$\eps^{\dot{\mu}\dot{\nu}}$, one obtains
\begin{equation}\label{Bkey}
-2\nabla_{\ald\bed\gad\ded}\hat{\CA}_{[\dot{1}\dot{2}]\dot{\rho}\dot{\sigma}}=
\CF_{\bed\gad\ded\ald\dot{\rho}\dot{\sigma}}~.
\end{equation}

The transversal gauge condition $\eta^\ald_k\hat{\CA}^k_\ald=0$ is
on $\CO_{(1;2,5)}$ equivalent to the condition
\begin{equation}\label{gauge2}
y^{(\ald\bed\gad\ded)}\hat{\CA}_{\ald\bed\gad\ded}=0~,
\end{equation}
as expected analogously to the case $\SA$. To lowest order in
$y^{(\ald\bed\gad\ded)}$, $\CF_{\alpha\beta}$ can be identified
with the self-dual field strength $f_{\alpha\beta}$ and
$\hat{\CA}_{[\dot{1}\dot{2}]\ald\bed}$ with the auxiliary field
$G_{\ald\bed}$. The remaining components of
$\CF_{\bed\gad\ded\ald\dot{\rho}\dot{\sigma}}$, i.e.\ those
antisymmetric in $[\ald\bed]$, are composite fields and do not
contain any additional degrees of freedom which is easily seen by
considering equation \eqref{Bkey}.

Applying the Euler operator in transverse gauge
$D:=y^{(\ald\bed\gad\ded)}\nabla_{(\ald\bed\gad\ded)}=
y^{(\ald\bed\gad\ded)}\dpar_{(\ald\bed\gad\ded)}$, one can show
that the lowest order field equations are equivalent to the full
superfield equations of motion. Thus, \eqref{compconB} is
equivalent to
\begin{equation}
f_{\ald\bed}=0\eand\nabla^{\alpha\ald}G_{\ald\bed}=0~.
\end{equation}
Altogether, we found the compatibility condition for a linear
system encoding purely bosonic SDYM theory {\em including} the
auxiliary field $G_{\ald\bed}$.

\section{Fattened real manifolds}

The field content of hCS theory on $\SA$ and $\SB$ becomes even
more transparent after imposing a reality condition on these
spaces. One can directly derive appropriate real structures from
the one on $\CP^{3|4}$, having in mind the picture of combining
the Gra{\ss}mann coordinates of $\CP^{3|4}$ to the even nilpotent
coordinates of $\SA$ and $\SB$. The real structure on $\CP^{3|4}$
is discussed in detail in \cite{Popov:2004rb}, one should note,
however, that our normalization of fields is different from the
one in this reference.

We have the following action of the two antilinear involutions
$\tau_\eps$ with $\eps=\pm 1$ on coordinates:
\begin{equation}\nonumber
\tau_\eps(z_+^1,z_+^2,\lambda_+)=\left(\frac{\bz_+^2}{\bl_+},
\frac{\eps\bz_+^1}{\bl_+},\frac{\eps}{\bl_+}\right)\eand
\tau_\eps(z_-^1,z_-^2,\lambda_-)=\left(\frac{\eps\bz_-^2}{\bl_-}
,\frac{\bz_-^1}{\bl_-},\frac{\eps}{\bl_-}\right)~.
\end{equation}
On $\SA$, we have additionally
\begin{equation}
\tau_\eps(y^1_+,y^2_+)=\left(\frac{\bar{y}^1_+}{\bl_+^2},\frac{\bar{y}^2_+}{\bl_+^2}\right)\eand
\tau_\eps(y^1_-,y^2_-)=\left(\frac{\bar{y}^1_-}{\bl_-^2},\frac{\bar{y}^2_-}{\bl_-^2}\right)~,
\end{equation}
and on $\SB$, it is
\begin{equation}
\tau_\eps(y_+)=\frac{\bar{y}_+}{\bl_+^4}\eand
\tau_\eps(y_-)=\frac{\bar{y}_-}{\bl_-^4}~.
\end{equation}
In the formulation of the twistor correspondence, the coordinates
$\lambda_\pm$ are usually kept complex for convenience sake
\cite{Popov:2004rb}. We do the same while on all other
coordinates, we impose the condition $\tau_{\eps}(\cdot)=\cdot$.
On the body of the moduli space, this will lead to a Euclidean
metric $(+,+,+,+)$ for $\eps=-1$ and a Kleinian metric $(+,+,-,-)$
for $\eps=+1$. Let us furthermore introduce the auxiliary
functions\footnote{One should note that the $\gamma_\pm$ are not
well-defined for $\eps=1$ on the whole of $\CP^3$, but only on the
subset for which $|\lambda|\neq 1$. Nevertheless, all the
formul\ae{} of the twistor correspondence can be used regardless
of this fact. Therefore we ignore this subtlety in the following
and refer to the discussion in
\cite{Popov:2004rb,Popov:2004nk,Lechtenfeld:2004cc}.}
\begin{equation}
\gamma_+=\frac{1}{1-\eps\lambda_+\bl_+}\eand
\gamma_-=-\eps\frac{1}{1-\eps\lambda_-\bl_-}~,
\end{equation}
and the notation
$(\hat{\lambda}^\ald)=(-\bl_{\dot{1}},\eps\bl_{\dot{2}})^T$.

The reality condition allows for the following identification:
\begin{equation}
\der{\bar{z}^1_+}=\gamma_+ V_2^+\eand\der{\bar{z}^2_+}=\eps
\gamma_+ V_1^+~,
\end{equation}
after which we can rewrite the hCS equations of motion, e.g.\ on
$\CU_+$, as
\begin{eqnarray}\label{shCS1}
V_\alpha^+\hat{\CA}_\beta^+- V_\beta^+\hat{\CA}_\alpha^++
[\hat{\CA}_\alpha^+,\hat{\CA}_\beta^+]&=&0~,\\
\label{shCS2} \dpar_{\bl_+}\hat{\CA}_\alpha^+-
V_\alpha^+\hat{\CA}_{\bl_+}+
[\hat{\CA}_{\bl_+},\hat{\CA}_\alpha^+]&=&0~,
\end{eqnarray}
where the components of the gauge potential are defined via the
contractions $\hat{\CA}^\pm_\alpha:=V_\alpha^\pm \lrcorner\,
\CA^{0,1}$, $\hat{\CA}_{\bl_\pm}:=\dpar_{\bl_\pm} \lrcorner\,
\CA^{0,1}$, and we assumed a gauge for which
$\hat{\CA}^\pm_{i}:=\dpar_{\bar{y}_i^\pm} \lrcorner\,
\CA^{0,1}=0$. On the space $\SA$ together with the field
expansion\footnote{Note that this expansion is determined by the
geometry of $\SA$, cf.\ \cite{Popov:2004rb}.}
\begin{eqnarray}\label{expAaA}
\hat{\CA}_\alpha^+&=&\lambda_+^\ald\, A_{\alpha\ald}+
\gamma_+\,y^+_i\,\hat{\lambda}^\ald_+\, \phi_{\alpha \ald}^{i}+
\gamma_+^3\,y^+_1y^+_2\,\hat{\lambda}_+^\ald\hat{\lambda}_+^\bed\hat{\lambda}_+^{\dot{\gamma}}\,
G_{\alpha\ald\bed\dot{\gamma}}~,\\
\label{expAlA}
\hat{\CA}_{\bl_+}&=&\gamma_+^2\,y^+_i\,\phi^i-2\eps\,\gamma_+^4\,y^+_1y^+_2\,
\hat{\lambda}_+^\ald\hat{\lambda}_+^\bed\, G_{\ald\bed}~,
\end{eqnarray}
the system of equations \eqref{shCS1} and \eqref{shCS2} is
equivalent to \eqref{compconA}. Furthermore, one can identify
$\phi^i_{\alpha\ald}=-\frac{1}{2}\CF^i_{\alpha\ald}$ and
$G_{\alpha\ald\bed\gad}=\frac{1}{6}\nabla^{(1}_{\ald(\bed}\CF^{2)}_{\alpha\gad)}$.
On $\SB$, we can use
\begin{eqnarray}\label{expAaB}
\hat{\CA}_\alpha^+&=&\lambda_+^\ald\, A_{\alpha\ald}+
\gamma_+^3\,y^+\hat{\lambda}_+^\ald\hat{\lambda}_+^\bed\hat{\lambda}_+^{\dot{\gamma}}\,
G_{\alpha\ald\bed\dot{\gamma}}~,\\
\label{expAlB} \hat{\CA}_{\bl_+}&=&\gamma_+^4\,y^+\,
\hat{\lambda}_+^\ald\hat{\lambda}_+^\bed\, G_{\ald\bed}~.
\end{eqnarray}
to have \eqref{shCS1} and \eqref{shCS2} equivalent with
\eqref{compconB} and
$G_{\alpha\ald\bed\gad}=\frac{1}{6}\CF_{\alpha\ald\bed\gad}$.

For compactness of the discussion, we refrain from explicitly
writing down all the reality conditions imposed on the component
fields and refer again to \cite{Popov:2004rb} for further details.

One can reconstruct two action functionals, from which the
equations of motion for the two cases arise. With our field
normalizations, they read
\begin{align}
S_{\SA}=&\int\dd^4 x \tr\left(G^{\ald\bed}f_{\ald\bed}-\phi^{(1}
\square
\phi^{2)}\right)~,\\
S_{\SB}=&\int\dd^4 x \tr\left(G^{\ald\bed}f_{\ald\bed}\right)~.
\end{align}
The action $S_{\SB}$ has first been proposed in
\cite{Chalmers:1996rq}.

\section{Exotic supermanifolds and Yau's theorem}

One can push the formalism of exotic supermanifolds a little bit
further to proceed with an analysis similar to \cite{Rocek:2004bi}
and ask, whether (an appropriate extension of) Yau's theorem is
valid for fattened complex manifolds which are Calabi-Yau, as the
ones considered in this paper\footnote{For related work, see
\cite{Zhou:2004su} and \cite{Rocek:2004ha}.}. This theorem states
that every K\"{a}hler manifold with vanishing first Chern class, or
equivalently, with a globally defined holomorphic volume form,
admits a Ricci-flat metric in every K\"{a}hler class.

We start from a $(k\oplus l|q)$-dimensional exotic supermanifold
with local coordinate vector
$(x^1,...,x^k,y^1,...,y^l,\zeta^1,...,\zeta^q)^T$. An element of
the tangent space is described by a vector
$(X^1,...,X^k,Y^1,...,Y^l,Z^1,...,Z^q)^T$. Both the metric and
linear coordinate transformations on this space are defined by
nonsingular matrices
\begin{equation}
K= \left(\begin{array}{ccc} A & B & C\\ D & E & F\\ G & H & J
\end{array}\right)~,
\end{equation}
where the elements $A,B,D,E,J$ are of even and $G,H,C,F$ are of
odd parity. As a definition for the {\em extended supertrace} of
such matrices, we choose
\begin{equation}
\etr(K):=\tr(A)+\tr(E)-\tr(J)~,
\end{equation}
which is closely related to the supertrace and which is the
appropriate choice to preserve cyclicity: $\etr(KM)=\etr(MK)$.
Similarly to \cite{DeWitt:1992cy}, we define the extended
superdeterminant by
\begin{equation}
\delta\ln\edet(K):=\etr(K^{-1}\delta K)~~~\mbox{together
with}~~~\edet(\unit):=1~,
\end{equation}
which guarantees $\edet(KM)=\edet(K)\edet(M)$. Proceeding
analogously to \cite{DeWitt:1992cy}, one decomposes $K$ into the
product of a lower triangular matrix, a block diagonal matrix and
an upper diagonal matrix. The triangular matrices can be chosen to
have only 1 as diagonal entries and thus do not contribute to the
total determinant. The block diagonal matrix is of the form
\begin{equation}
K'=\left(\begin{array}{ccc} A & 0 & 0\\
0 & E-DA^{-1}B & 0 \\ 0 & 0 & R
\end{array}\right)~,
\end{equation}
with $R=J-GA^{-1}C-(H-GA^{-1}B)(E-DA^{-1}B)^{-1}(F-DA^{-1}C)$. The
determinant of a block diagonal matrix is easily calculated and in
this case we obtain
\begin{equation}
\edet(K)=\edet(K')=\frac{\det(A)\det(E-DA^{-1}B)}{\det(R)}~.
\end{equation}
Note that for the special case of no even nilpotent dimensions,
for which one should formally set $B=D=F=H=0$, one recovers the
formul\ae{} for the supertrace ($E=0$) and the superdeterminant
($E=\unit$ to drop the additional determinant).

In \cite{Rocek:2004bi}, the authors found that K\"{a}hler
supermanifolds with one fermionic dimension admit Ricci-flat
supermetrics if and only if the body of the K\"{a}hler supermanifold
admits a metric with vanishing scalar curvature. Let us
investigate the same issue for the case of an $(p\oplus
1|0)$-dimensional exotic supermanifold $Y$ with one even nilpotent
coordinate $y$. We denote the ordinary $p$-dimensional complex
manifold embedded in $Y$ by $X$. The extended K\"{a}hler potential on
$Y$ is given by a real-valued function $\CK=f^0+f^1y\bar{y}$, such
that the metric takes the form
\begin{equation}
g:=\left(\dpar_i\dparb_{\bar{\jmath}}\CK\right)=\left(\begin{array}{cc}
f^0_{,i\bar{\jmath}}+f^1_{,i\bar{\jmath}}\,y\bar{y} & f^1_{,i}\,y \\
f^1_{,\bar{\jmath}}\,\bar{y} & f^1
\end{array}\right)~.
\end{equation}
For the extended Ricci-tensor to vanish, the extended K\"{a}hler
potential has to satisfy the Monge-Amp{\`e}re equation
$\edet(g):=\edet(\dpar_i\dparb_{\bar{\jmath}}\CK)=1$. In fact, we
find
\begin{align*}
\edet(g)=&\det\left(f^0_{,i\bar{\jmath}}+f^1_{,i\bar{\jmath}}y\bar{y}\right)\left(f^1-
f^1_{\bar{m}}g^{\bar{m}n}f^1_{,n}y\bar{y}\right)\\
=&\det\left[\left(f^0_{,i\bar{\jmath}}+f^1_{,i\bar{\jmath}}y\bar{y}\right)\left(\sqrt[p]{f^1}-
\frac{f^1_{\bar{m}}g^{\bar{m}n}f^1_{,n}y\bar{y}}{p
(f^1)^{\frac{p-1}{p}}}\right)\right]\\
=&\det\left[f^0_{,i\bar{\jmath}}\sqrt[p]{f^1}+\left(f^1_{,i\bar{\jmath}}\sqrt[p]{f^1}-
f^0_{,i\bar{\jmath}}\frac{f^1_{\bar{m}}g^{\bar{m}n}f^1_{,n}}{p
(f^1)^{\frac{p-1}{p}}}\right)y\bar{y}
\right]\\
=&\det\left[f^0_{,l\bar{\jmath}}\sqrt[p]{f^1}\right]
\det\left[\delta_i^k+\left(g^{\bar{m}k}f^1_{,i\bar{m}}-
\delta_i^k\frac{f^1_{\bar{m}}g^{\bar{m}n}f^1_{,n}}{p
f^1}\right)y\bar{y} \right]~,
\end{align*}
where $g^{\bar{m}n}$ is the inverse of $f^0_{,n\bar{m}}$. Using
the relation $\ln\det(A)=\tr \ln(A)$, we obtain
\begin{equation}
\edet(g)=\det\left[f^0_{,l\bar{\jmath}}\sqrt[a]{f^1}\right]\left(1+\left(g^{\bar{m}i}f^1_{,i\bar{m}}-
\frac{f^1_{\bar{m}}g^{\bar{m}n}f^1_{,n}}{f^1}\right)y\bar{y}
\right)~.
\end{equation}
From demanding extended Ricci-flatness, it follows that
\begin{equation}\label{cond0}
f^1=\frac{1}{\det\left(f^0_{,l\bar{\jmath}}\right)}\eand
\left(g^{\bar{\jmath}i}f^1_{,i\bar{\jmath}}-
\frac{f^1_{\bar{\jmath}}g^{\bar{\jmath}i}f^1_{,i}}{
f^1}\right)=0~.
\end{equation}
The second equation can be simplified to
\begin{equation}
g^{\bar{\jmath}i}\left(f^1_{,i\bar{\jmath}}-
\frac{f^1_{\bar{\jmath}}f^1_{,i}}{ f^1}\right)= f^1
g^{\bar{\jmath}i}\left(\ln(f^1)\right)_{,i\bar{\jmath}}=0~,
\end{equation}
and together with the first equation in \eqref{cond0}, it yields
\begin{equation}
g^{\bar{\jmath}i}\left(\ln
\frac{1}{\det\left(f^0_{,k\bar{\jmath}}\right)}\right)_{,i\bar{\jmath}}=
-g^{\bar{\jmath}i}\left(\ln\det\left(f^0_{,k\bar{\jmath}}\right)\right)_{,i\bar{\jmath}}=
-g^{\bar{\jmath}i}R_{,i\bar{\jmath}}=0~.
\end{equation}
This equation states that an exotic supermanifold $Y$ of dimension
$(p\oplus 1|0)$ admits an extended Ricci-flat metric if and only
if the embedded ordinary manifold $X$ has vanishing scalar
curvature. A class of examples for which this additional condition
is not satisfied are the weighted projective spaces
$W\CPP^{m-1\oplus1|0}(1,...,1\oplus m|\cdot)$ which have vanishing
first Chern class but do not admit a K\"{a}hler metric with vanishing
Ricci scalar.

Thus, we obtained exactly the same result as in
\cite{Rocek:2004bi}, which is somewhat surprising as the
definition of the extended determinant involved in our calculation
strongly differs from the definition of the superdeterminant.
However, this agreement might be an indication that fattened
complex manifolds -- together with the definitions made in this
paper -- fit nicely in the whole picture of extended Calabi-Yau
spaces.

\section*{Acknowledgements}
I would like to thank Michael Eastwood and Albert S. Schwarz for
correspondence and helpful remarks on the definition of
integration on exotic supermanifolds. I am also very grateful to
Sebastian Uhlmann and Martin Wolf for reading and commenting in
detail a first draft. In particular, I want to express my
gratitude towards Alexander D. Popov for sharing his ideas and for
valuable comments which resulted in many improvements. This work
was done within the framework of the DFG priority program (SPP
1096) in string theory.

\end{document}